\documentclass[11pt]{article}
\pdfoutput=1
\usepackage{dcolumn}
\usepackage{bm}
\usepackage{graphicx}
\usepackage{amssymb,amsmath}
\usepackage{multirow}
\usepackage{cite,color,url}
\usepackage[colorlinks=true
,urlcolor=blue
,anchorcolor=blue
,citecolor=blue
,filecolor=blue
,linkcolor=blue
,menucolor=blue
,linktocpage=true
,pdfproducer=medialab
,pdfa=true
]{hyperref}

\usepackage{slashed}
\usepackage{epsfig,psfrag,rotating,soul}
\usepackage{rotfloat}
\usepackage[font={small}]{caption}
\usepackage{colortbl}
\usepackage{xcolor}

\evensidemargin \oddsidemargin
\allowdisplaybreaks

\let\OLDthebibliography\thebibliography
\renewcommand\thebibliography[1]{
  \OLDthebibliography{#1}
  \setlength{\parskip}{0pt}
  \setlength{\itemsep}{0pt plus 0.3ex}
}

\definecolor{mygray}{gray}{0.85} 
\definecolor{myblue}{cmyk}{0.65, 0.37, 0.0, 0.19}

\begin{document}
\thispagestyle{empty}

\def\thefootnote{\fnsymbol{footnote}}

\begin{flushright}
IFT-UAM/CSIC-21-90
\end{flushright}

\vspace*{1cm}

\begin{center}

\begin{Large}
\textbf{\textsc{Interpretation of LHC excesses in ditop and ditau channels \\[.25em] as a 400-GeV pseudoscalar resonance}}
\end{Large}

\vspace{1cm}

{\sc
Ernesto~Arganda$^{1, 2}$%
\footnote{{\tt \href{mailto:ernesto.arganda@csic.es}{ernesto.arganda@csic.es}}}%
, Leandro~Da~Rold$^{3}$%
\footnote{{\tt \href{mailto:daroldl@cab.cnea.gov.ar}{daroldl@cab.cnea.gov.ar}}}%
, Daniel~A.~D\'{\i}az$^{2}$%
\footnote{{\tt \href{mailto:daniel.diaz@fisica.unlp.edu.ar}{daniel.diaz@fisica.unlp.edu.ar}}}%
and Anibal D.~Medina$^{2}$%
\footnote{{\tt \href{mailto:anibal.medina@fisica.unlp.edu.ar}{anibal.medina@fisica.unlp.edu.ar}}}%

}

\vspace*{.7cm}

{\sl
$^1$Instituto de F\'{\i}sica Te\'orica UAM/CSIC, \\
C/ Nicol\'as Cabrera 13-15, Campus de Cantoblanco, 28049, Madrid, Spain

\vspace*{0.1cm}

$^2$IFLP, CONICET - Dpto. de F\'{\i}sica, Universidad Nacional de La Plata, \\ 
C.C. 67, 1900 La Plata, Argentina

\vspace*{0.1cm}

$^3$Centro At\'omico Bariloche, Instituto Balseiro and CONICET, \\
Av. Bustillo 9500, 8400, S. C. de Bariloche, Argentina
}

\end{center}

\vspace{0.1cm}

\begin{abstract}
\noindent
Since the discovery in 2012 of the Higgs boson at the LHC, as the last missing piece of the Standard Model of particle physics, any hint of new physics has been intensively searched for, with no confirmation to date. There are however slight deviations from the SM that are worth investigating. The CMS collaboration has reported, in a search for heavy resonances decaying in $t \bar t$ with a 13-TeV center-of-mass energy and a luminosity of 35.9 fb$^{-1}$, deviations from the SM predictions at the 3.5$\sigma$ level locally (1.9$\sigma$ after the look-elsewhere effect). In addition, in the ditau final state search performed by the ATLAS collaboration at $\sqrt{s}=13$ TeV and $\mathcal{L}=139$ fb$^{-1}$, deviations from the SM at the 2$\sigma$ level have been also observed. Interestingly, both slight excesses are compatible with a new pseudoscalar boson with a mass around 400 GeV that couples at least to fermions of the third generation and gluons. Starting from a purely phenomenological perspective, we inspect the possibility that a 400-GeV pseudoscalar can account for these deviations and at the same time satisfy the constraints on the rest of the channels that it gives contributions to and that are analyzed by the ATLAS and CMS experiments. After obtaining the range of effective couplings compatible with all experimental measurements, we study the gauge invariant UV completions that can give rise to this type of pseudoscalar resonance, which can be accommodated in an $SO(6)/SO(5)$ model with consistency at the 1$\sigma$ level and in a $SO(5)\times U(1)_{P}\times U(1)_{X}/SO(4)\times U(1)_X$ at the 2$\sigma$ level, while exceedingly large quartic couplings would be necessary to account for it in a general two Higgs doublet model.
\end{abstract}

\def\thefootnote{\arabic{footnote}}
\setcounter{page}{0}
\setcounter{footnote}{0}

\newpage

\section{Introduction}
\label{intro}

There are many searches for new physics being done at the Large Hadron Collider (LHC) but for the moment only a scalar particle that resembles the last missing piece of the Standard Model (SM) has been discovered, the Higgs~\cite{Aad:2012tfa,Chatrchyan:2012ufa}. The fact that no clear indication of signs of new physics emerges from these collider searches implies that, as phenomenologists, we should pay attention to the small but coordinated deviations from the SM behavior that may arise in different search channels as possible hints of a specific type of new physics. Interestingly, there are two particular somewhat recent searches in which this behavior seems to be happening. On one hand, in the ditop search performed by the CMS collaboration~\cite{Sirunyan:2019wph} at a center of mass energy  $\sqrt{s}=13$ TeV and a luminosity of $\mathcal{L}=35.9$ fb$^{-1}$, with top quarks decaying into single and dilepton final states, deviations from the SM behavior arise at the 3.5 $\sigma$ level locally (1.9 $\sigma$ after the look-elsewhere effect) that can be accounted for by a pseudoscalar with a mass around 400 GeV that couples at least to top quarks and gluons. On the other hand, in the ditau search done by the ATLAS collaboration~\cite{Aad:2020zxo} at $\sqrt{s}=13$ TeV and $\mathcal{L}=139$ fb$^{-1}$, deviations from the SM at the 2$\sigma$ level are observed which, interestingly enough, can be accounted for by pseudoscalar coupling to tau leptons, bottom-quarks and gluons,  once again with a mass around 400 GeV. Furthermore, whatever new physics may account simultaneously for these possible hints should also be consistent with other collider searches to which it could provide contributions to~\cite{Aad:2014ioa,Khachatryan:2015qba,Aaboud:2017yyg,Aaboud:2017uhw,Aaboud:2017cxo,Sirunyan:2018taj,Sirunyan:2019xls,Sirunyan:2019xjg,Sirunyan:2019wrn,Aad:2020klt}.

In this work we consider initially from a pure phenomenological bottom-up approach the possibility that a pseudoscalar with a mass of 400 GeV is indeed responsible simultaneously for these observed hints and at the same time satisfies the constraints coming from other final states that it could contribute to and that are search for at the LHC. An idea in this direction was considered also in~\cite{Richard:2020cav}. Considering the interactions of the pseudoscalar up to dimension-5 operators, we find that there exists a well-defined region of couplings in which the hints as well as the constraints can be satisfied at the same time. We then study the implications on the new physics of possible gauge invariant models in which a pseudoscalar  with that particular mass and couplings can be obtained. We find that, given the parameter space consistent with the measurements, it turns out to be highly unlikely that the pseudoscalar could be accommodated in a two Higgs doublet model (2HDM) even in more general possible versions of it~\cite{Egana-Ugrinovic:2019dqu}. However, we show that if the pseudoscalar is associated with a pseudo-Nambu-Goldstone (pNGB) boson of broken global symmetry in composite Higgs models~\cite{Kaplan:1983fs,Kaplan:1983sm,Georgi:1984ef,Georgi:1984af,Dugan:1984hq,Contino:2003ve,Agashe:2004rs}, one can accommodate its mass and couplings for an $SO(6)/SO(5)$ model~\cite{Gripaios:2009pe,Chala:2017sjk} with consistency and the 1$\sigma$ level with the measurements, and in a $SO(5)\times U(1)_{P}\times U(1)_{X}/SO(4)\times U(1)_X$~\cite{Chala:2017sjk} at the 2$\sigma$ level.

The remainder of the work is organized as follows: in Section~\ref{exp-searches} we summarize the main experimental searches, carried out by ATLAS and CMS, for heavy resonances that give rise to small deviations from the SM expectations or that could restrict the possible parameter space. Section~\ref{pheno} is devoted, from a general approach of effective field theories, to the phenomenology of a 400-GeV pseudoscalar boson that can account for these experimental signatures. In Section~\ref{num-results} we analyze numerically the range of couplings allowed and excluded by the experimental data, while Section~\ref{UVcomp} is dedicated to the UV completions that could give rise to this 400-GeV pseudoscalar boson with the allowed couplings. Finally, we present our main conclusions in Section~\ref{conclu}.

\section{Experimental hints and constraints}
\label{exp-searches}
In the present work we consider several searches carried out at the LHC by the ATLAS and CMS collaborations that either show a hint or lead to a constraint for a pseudoscalar, and that allow us to determine the effective couplings of the proposed new state. A brief summary of the searches and the parameters used is presented in this section.

\subsection{Final state $t\bar t$} 
A moderate excess in top quark pair production has been found by the CMS collaboration \cite{Sirunyan:2019wph}. This excess~\footnote{We would like to stress that  higher-order electroweak corrections to the SM-$t \bar t$ production can become important in the vicinity of the pair production threshold~\cite{Kuhn:2013zoa} and may be responsible for the excess observed. It is therefore important that further improvements of the theoretical description of the standard model $t \bar t$ process in the vicinity of the production threshold are calculated in order to clarify the origin of this deviation.} has been proved to be compatible with an intermediate state consisting on a scalar or pseudoscalar boson with a mass of $400$ GeV created via gluon fusion and decaying into a top-antitop quark pair. In Ref.~\cite{Sirunyan:2019wph} the CMS collaboration presented a search for heavy Higgs bosons decaying into a top quark pair in single and dilepton final states within a data set corresponding to an integrated luminosity of 35.9 fb$^{-1}$, with a center-of-mass energy of $13$ TeV. The masses of the hypothetical scalar and pseudoscalar bosons were probed within the range $400$ to $750$ GeV and a total relative width from $0.5$ to $25\%$ of its mass. The largest deviation from the SM background was observed for a pseudoscalar boson with a mass of $400$ GeV and a total relative width of $4\%$, with a local significance of $3.5\pm 0.3$ standard deviations. The significance of the excess becomes $1.9$ standard deviations after accounting for the look-elsewhere effect in the mass, total width and CP-state of the new resonance. The analysis considered tree level interactions with the top quark only, as well as interactions with gluons induced at one-loop level. Under those assumptions it was found that, for $m_A=400$ GeV and $\Gamma_A/m_A=0.04$, a top coupling $g_{At\bar t}\approx 0.9$ yields the maximum likelihood ratio between the hypothesis of the existence of the pseudoscalar boson and the SM scenario, as can be seen in Fig. 7 of that work.~\footnote{We are using the same notation as Ref.~\cite{Sirunyan:2019wph} in this subsection.} For $0.6\lesssim g_{At\bar t}\lesssim 1$ the model exceeds the range of compatibility with the SM in more than two standard deviations, is consistent with the measurement at one standard deviation and is below a critical value which would yield a nonphysical scenario, in which the partial decay width of the pseudoscalar boson into a top-antitop pair would exceed the total decay width from that channel.
In the present work we consider cross sections of the $t\bar t$ channel in agreement with those values of $g_{At\bar t}$.

 There is no current 13-TeV ATLAS search in the ditop channel though there is a search at 8 TeV with 20.3 fb$^{-1}$~\cite{ATLAS:2017snw}. In final states with an electron or muon, large missing transverse momentum, and at least four jets, there is no significant deviation from the SM prediction with the caveat that dileptonic final states and angular variables were not utilized for this analysis, and cross section limits were set only for masses from 500 GeV to 750 GeV. Therefore a combined analysis is not possible.

\subsection{Final state $\tau \tau$} 
In Ref.~\cite{Aad:2020zxo} the ATLAS collaboration reported a search for heavy scalar and pseudoscalar bosons performed with data corresponding to Run 2 of the LHC, with an integrated luminosity of $139$ fb$^{-1}$ at a center-of-mass energy of $\sqrt{s}=13$ TeV. The search for heavy resonances was performed within the mass range $0.2-2.5$ TeV, in the $\tau^+\tau^-$ decay channel. The relevant data for this work is presented in plots of $\sigma_{bb}\times BR_{\tau \tau}$ vs $\sigma_{gg}\times BR_{\tau \tau}$, that show ellipses for one and two standard deviations containing the observed value at the center. For a mass of 400 GeV the SM scenario lies more than two standard deviations away from the observed value. In this work we demand $\sigma \times BR$ lying within the area contended by the $1\sigma$ ellipses (excepted where explicitly stated).

The corresponding CMS ditau final state search was done with  an integrated luminosity of 35.9 fb $^{-1}$~\cite{CMS:2018rmh}. Similarly to the ATLAS search, gluon  and bottom fusion production modes were considered, searching for heavy resonances in the mass range 90 GeV - 3.2 TeV. No excess over the SM background was observed. Nonetheless,  the expected and observed cross section limits from the CMS analysis are weaker than the ones from ATLAS given the smaller amount of data included in comparison to the ATLAS search.  This implies that the signal interpretation of the ATLAS excess at around 400 GeV  is compatible with the expected (observed) upper limits from the CMS analysis at the level of 1.1(1.9)$\sigma$~\cite{Biekotter:2021qbc}.
 
\subsection{Final state $b\bar b$} 
The CMS collaboration has searched for Higgs bosons decaying into a bottom-antibottom quark pair accompanied by at least one additional bottom-quark, with data corresponding to a center-of-mass energy of $13$ TeV and an integrated luminosity of 35.7 fb$^{-1}$,~\cite{Sirunyan:2018taj}. The analysis considered scalar and pseudoscalar bosons with masses ranging from $300$ to $1300$ GeV, finding no significant deviation from the SM. An upper limit for the cross section times branching fraction is reported: $\sigma_{bb} \times BR_{bb}=5.7$ pb at 95\% confidence level (CL), for a mass of 400 GeV.

\subsection{Final state $t\bar tt\bar t$} 
Another decay channel of interest for the present work consists on the decay of the pseudoscalar boson into four tops. We consider the result reported in Ref.~\cite{Aad:2020klt} by the ATLAS collaboration, where a production cross section of two pairs top-antitop was found to be $24^{+7}_{-6}$ fb. Interestingly, notice that this value is roughly two standard deviations above the SM prediction.

\subsection{Final state $Zh$}
A pseudoscalar boson decaying into a $Z$ boson and a neutral SM Higgs boson has been probed by the ATLAS and CMS collaborations in multiple searches, finding no evidence of any significant deviation from the SM background. In the present work we consider the constraints for $\sigma\times BR$ reported by the ATLAS collaboration~\cite{Aaboud:2017cxo}, that sets limits in the gluon fusion and bottom fusion initial states, for a mass of 400 GeV the 95\% CL bounds are roughly $0.22$ pb and $0.25$ pb, respectively. 

\subsection{Final state $\gamma\gamma$} 
Searches for new heavy particles decaying into two photons within mass ranges containing $400$ GeV have been carried out by the ATLAS and CMS collaborations. No excess has been found by any search and upper limits were set for production cross section times branching ratio.

In Ref.~\cite{Aad:2014ioa} the ATLAS collaboration presented a search for scalar particles decaying via narrow resonances into two photons with masses ranging $65$ to $600$ GeV, using $20.3$ fb$^{-1}$ at $\sqrt{s}=8$ TeV, having found no evidence for the existence of these particles and setting an upper limit of $\sigma\times BR_{\gamma\gamma}\lesssim 2-3$ fb at 95\% CL. Another search performed by the ATLAS collaboration is presented in Ref.~\cite{Aaboud:2017yyg} with an integrated luminosity of $36.7$ fb$^{-1}$ at $\sqrt{s}=13$ TeV. This search included spin-0 particles decaying into a final state consisting on two photons within a mass range of $200$ to $2700$ GeV. For 400 GeV the upper limit was found to be $2-3$ fb at 95\% CL. In Ref.~\cite{Khachatryan:2015qba} the CMS collaboration explored the diphoton mass spectrum from $150$ to $850$ GeV with an integrated luminosity of $19.7$ fb$^{-1}$, at a center of mas energy of $\sqrt{s}=8$ TeV. Assuming a total decay width between $0.1$~GeV and $40$~GeV the CMS collaboration reported an upper limit $\lesssim 4-15$~fb at 95\% CL.

\subsection{Final state $VV$}
The $VV$ final state, with $V=W,Z$, has been explored by the ATLAS and CMS collaborations at the LHC in numerous searches, finding no excesses in any of them. 

A search for heavy neutral resonances decaying into a $W$ boson pair was carried out by the ATLAS collaboration in Ref.~\cite{ATLAS:2017uhp} using a data set corresponding to an integrated luminosity of 36.1 fb$^{-1}$ with a center-of-mass energy of $13$ TeV. The search focuses on the decay channel $(WW\rightarrow e\nu\mu\nu)$ and provides an upper limit for $\sigma\times BR_{WW}$ as a function of the mass of the resonance, ranging from $200$ GeV to $5$ TeV. For the analysis various benchmark models were considered, including Higgs-like scalars in different width scenarios. 
In Ref.~\cite{ATLAS:2020tlo}, the ATLAS collaboration presents a search for heavy resonances decaying into a pair of $Z$ bosons leading to two pairs lepton-antilepton, or two pairs lepton-neutrino, for masses ranging from $200$ to $2000$ GeV. A similar search was published by the CMS collaboration in the range $130$ GeV to $3$ TeV~\cite{CMS:2018amk}.

\subsection{Final state $Z\gamma$} 

A search for the Higgs boson and for narrow high mass resonances decaying into $Z\gamma$ is presented by the ATLAS collaboration in Ref.~\cite{Aaboud:2017uhw} using $36.1$ fb$^{-1}$ of $pp$ collisions at $\sqrt{s}=13$ TeV. The search for high mass resonances focuses on spin-0 and spin-2 interpretations. The results are found to be consistent with the SM and upper limits on the production cross section times the branching ratio are reported, varying the observed values between $88$ fb and $2.8$ fb for masses within $250$-$2400$ GeV in the case of spin-0 resonances.

\section{Phenomenology}
\label{pheno}

Considering the experimental hints described in the previous section as a possible sign of new physics at an invariant mass around 400 GeV, we propose initially from a purely phenomenological perspective to do an analysis in which we study the compatibility of some of these hints with the introduction of a new pseudoscalar state $a$ with a mass $m_a=400$ GeV, that interacts solely with $3^{rd}$ generation charged quarks and leptons, the SM gauge bosons $Z$, gluons, photons, and $H$, as:
\begin{align}\label{eq-Lint}
{\cal L}_{\rm int}&=\frac{g_{gg}}{4}\ a\ G_{\mu\nu}\tilde G^{\mu\nu} + \frac{g_{\gamma\gamma}}{4}\ a\ F_{\mu\nu}\tilde F^{\mu\nu} + g_{Zh} (a\overleftrightarrow\partial_\mu h) Z^\mu
 + i g_{t}\ a\ \bar t\gamma^5 t 
+  i g_{b}\ a\ \bar b\gamma^5 b + i g_{\tau}\ a\ \bar\tau\gamma^5\tau \ ,
\end{align}
with $\tilde F^{\mu\nu}=(1/2)\epsilon^{\mu\nu\rho\sigma}F_{\rho\sigma}$ and similarly for $\tilde G^{\mu\nu}$, $g_{t}$, $g_{b}$, $g_{\tau}$, and $g_{Zh}$ dimensionless couplings, whereas $g_{gg}$ and $g_{\gamma\gamma}$ have dimension of an inverse mass scale, the first ones are expected to be present at tree level, whereas the second ones are expected to be induced at the one-loop level. This is the most general CP-invariant interaction Lagrangian linear in the pseudoscalar state $a$ that can be written up to dimension-5 operators, neglecting pseudoscalar interactions with the kinetic terms of massive electroweak gauge bosons $\tilde{Z}_{\mu\nu}Z^{\mu\nu}$ and $\tilde{W}^{\dagger}_{\mu\nu}W^{\mu\nu}$ and with $\tilde{F}_{\mu\nu}Z^{\mu\nu}$ that, though they are of the same order as the interaction with photons (1-loop order), for the phenomenology we want to address they turn out to be irrelevant. In fact, due to their connection via electroweak (EW) symmetry, the pseudoscalar coupling to EW massive gauge bosons and to $Z\gamma$ should be similar in nature to the diphoton coupling. Re-scaling the photon coupling by the corresponding factors and modifying the decay rates by the appropriate phase spaces we obtain, in the interesting region of couplings, $VV$ and $Z\gamma$ cross sections that are one to two orders of magnitude below the bounds discussed in the previous section.

In order to be able to separate the contribution of possible heavy colored and/or electromagnetically charged beyond the SM (BSM) states to $g_{gg}$ and $g_{\gamma\gamma}$ that have been integrated out from our effective theory,  from those of the $3^{rd}$ generation quarks and leptons, we explicitly write  $g_{gg}$ and $g_{\gamma\gamma}$ as,
\begin{eqnarray}
g_{gg}=\frac{3\alpha_s}{12\pi m_t}\times 4.2 (x+ g_{t})\; , \quad g_{\gamma\gamma}=\frac{3\alpha}{2\pi m_t}\times \left(\frac{2}{3}\right)^2 4.2 \left(z+ g_{t}+ x\left(\frac{Q_x}{2/3}\right)^2\right) \ ,
\end{eqnarray}
with $\alpha_s=g^2_s/(4\pi)$ and $\alpha= e^2/(4\pi)$ and where we are already assuming that in Eq.~(\ref{eq-Lint}), in what concerns the SM quark contributions, only the top quark provides sizable contributions to the loop-induced couplings and we explicitly used that the loop function $A^{a}_{1/2}(m_a^2/(4m_t^2))\approx 4.2$ for $m_a=400$ GeV. Furthermore, we parameterize the contributions from heavy BSM states that have been integrated out as coming from vector-like fermions in the fundamental representation of $SU(3)_c$ with EM charge $Q_x$, whose contribution is accounted by the dimensionless parameter $x$ and a separate BSM contribution from colorless vector-like fermions charged under EM and accounted by the dimensionless parameter $z$. These loop contributions are normalized with respect to the top quark contribution such that, for example, in the presence of a color triplet heavy fermion $F$ with EM charge $Q_x=Q_F$, are:
\begin{align}
x=x_F= 
\frac{g_{F}\ m_t}{m_F}\times \frac{2}{4.2}  \ ,
\end{align}
where we used $A^a_{1/2}(m_a^2/(4m_F^2))\approx 2$ for $m_F\gg m_a$, and $g_{F}$ is the vector-like coupling of $a$ to the heavy fermion $F$. Analogously, for a colorless vector-like fermion $E$ with EM charge $Q_E$ we would obtain,
\begin{align}
z=z_E=
\frac{1}{3}\left(\frac{Q_E}{2/3}\right)^2 \frac{g_{E}\ m_t}{m_E}\times \frac{2}{4.2}  \ .
\end{align}

We restrict ourselves to the case in which the pseudoscalar behaves as a narrow resonance, narrow width approximation (NWA),  with a total width $\Gamma_a/m_a\leq 8 \%$. The dominant 2-body partial decay widths for $a$ take the form,
\begin{eqnarray}
\Gamma_{a\to f\bar{f}}&=&N_c\frac{g^2_{f}m_a}{8\pi}\sqrt{1-4\frac{m^2_f}{m^2_a}} \ ,\\
\Gamma_{a\to Zh}&=&\frac{g^2_{Zh}}{16\pi }\frac{m^3_a}{m^2_Z}\left[1+\left(\frac{m^2_h-m^2_Z}{m^2_a}\right)^2-2\left(\frac{m^2_h+m^2_Z}{m^2_a}\right)\right]^{3/2} \ ,\\
\Gamma_{a\to gg}&=&\frac{  g^2_{gg }m^3_a}{8\pi} \ ,\\
\Gamma_{a\to \gamma\gamma}&=&\frac{ g^2_{\gamma\gamma}m^3_a }{8\pi} \ ,
\end{eqnarray}
where $f$ represents the fermions of interest, $f=t,b,\tau$ and $N_c=3$ for $f=t,b$ and $N_c=1$ for $\tau$.

In what respects the production of the pseudoscalar, the main channels are gluon and bottom fusion. Following Ref.~\cite{Franceschini:2015kwy}, the hadronic production cross section from parton pairs ${\cal P}\bar{\cal P}$ at an energy $\sqrt{s}$ can be written as,
\begin{equation}
\sigma(pp\to a) = \frac{1}{s\times m_a}\sum_{\cal P} C_{{\cal P}\bar{\cal P}}\Gamma_{a\to{\cal P}\bar{\cal P}} \ .
\end{equation}
For the LHC at $\sqrt{s}=8,\,13$ TeV, considering only couplings to gluons and bottom-quarks, we obtain the following:
\begin{center}
\begin{tabular}{| c | c | c |}
\hline
$\sqrt{s}$ & $C_{gg}$ & $C_{b\bar b}$
\\ \hline
8 TeV & 4592 & 29
\\ \hline
13 TeV & 40187 & 278
\\ \hline
\end{tabular}
\end{center}
The typical values that we find for the production cross sections at the LHC for $\sqrt{s}=13$ TeV that are consistent with the experimental hints are $\sim 20$ pb and $\sim 10$ pb for gluon and bottom fusion,  respectively.

Having all these elements at hand, we can calculate under the NWA, the contribution of $a$ to the different final states that either show a possible hint of BSM physics or otherwise put a constraint on the couplings of the pseudoscalar to SM particles. We do this in the following section and comment on our findings. In particular we focus on the cases in which there could be a UV contribution to the gluonic operator $x\neq 0$ and when such contribution vanishes, $x=0$.

\section{Numerical results}
\label{num-results}

In this section we study the cross sections of the different production modes of the 400-GeV pseudoscalar, taking into account their dominant decay channels, as a function of two parameters at a time, fixing the rest of the parameters to two benchmark points, one with $x$ = 0 and another one with $x$ $\neq$ 0, defined as follows:
\begin{itemize}
\item $g_t$ = 0.78, $g_b$ = 0.39, $g_\tau$ = 0.04, $x$ = 0,
\item $g_t$ = 0.7, $g_b$ = 0.37, $g_\tau$ = 0.04, $x$ = 0.13.
\end{itemize}
All the cross sections have been computed with the expressions detailed in the previous section, except the $t \bar t t \bar t$ production, which has been calculated with {\tt MadGraph\_aMC@NLO 2.5.2}~\cite{Alwall:2014hca} at leading order. For the pseudoscalar productions via gluon fusion and in association with bottom-quarks we have considered $K$-factors of 2~\cite{Sirunyan:2019wph} and 1.24~\cite{Bonvini:2016fgf}, respectively, while a $K$-factor of 1.27~\cite{Aad:2020klt}, with a 20\% uncertainty, has been used for $t \bar t t \bar t$ production.
Likewise, for each final cross section we consider 1$\sigma$ uncertainties, also detailed in Section~\ref{exp-searches}. We present the results as contour plots of the different cross sections as a function of a pair of couplings, and each contour line is labeled with the corresponding cross-section value at $\pm$1$\sigma$. The red shaded areas correspond to values of the cross sections lower than their reference values subtracting their uncertainty of 1$\sigma$, and the blue ones to higher than these reference values plus 1$\sigma$. In the case of the ditau channel, we demand consistency at the 1$\sigma$ level with the experimental measurements  (given by an ellipse in the gluon and bottom fusion channels, see Section~\ref{exp-searches}). Therefore, both blue and red shaded areas are excluded by data, and only the white area would be allowed by the LHC experimental measurements and consistent with the possible hints of new physics. Recall also that the only channels that show slight excesses or deviations from the SM predictions are $t \bar t$ and $\tau^+ \tau^-$, as well as $t \bar t t \bar t$, while the other channels only impose exclusion limits on the total cross sections at 95\% CL. In addition, it is important to note that the channels that are not shown in the following contour plots are due to the fact that they impose softer constraints on our parameter space. In particular, we find that the coupling to a $Z$ and a Higgs is restricted to: $g_{Zh}\lesssim 0.03$, roughly independently of the other couplings, while the new physics contribution to diphotons, called $z$, shows a stronger dependence on $g_t$ and $x$, with larger values allowed for smaller $g_t$ and $x$: $z\lesssim 1.2$ for $g_t=0.7$ and $x=0$, and smaller values for larger $g_t$ and $x$: $z\lesssim 0.5$ for $g_t=0.78$ and $x=0.13$. Neither $g_{Zh}$ nor $z$ determine the shape of the white regions in the plots we show.

Concerning the branching ratios of the dominant decay channels of the 400-GeV pseudoscalar, we find that for the parameter region consistent with the experimental measurements, they are dominated by either ditops of dibottoms since we find that the pseudoscalar coupling to ditops is large, but more interestingly the consistency with measurements implies a large coupling to dibottoms, up to 30 times the Higgs bottom-quark Yukawa coupling. For the benchmark scenario with $x$ = 0, the maximum value for BR($a \to t \bar t$) allowed by our analysis is 0.970, the minimum one is 0.270 and the average value is 0.675, while the maximum value for BR($a \to b \bar b$) is 0.726, the minimum one is 0.022 and the average value is 0.319. On the other hand, within the benchmark scenario with $x$ = 0.13, the maximum, minimum, and average values allowed by our analysis for the $t \bar t$ ($b \bar b$) decay modes are 0.806, 0.541, and 0.696 (0.450, 0.187, and 0.296), respectively. Notice also that the contribution of the bottom-quark loop to the production of the 400-GeV pseudoscalar via gluon fusion, for the favored values of the $g_b$ and $g_t$ couplings, is of order 6-8\% compared to that of the top-quark loop.~\footnote{Also as comparison, for the production of the SM Higgs boson, the $b$-quark contribution to the gluon-fusion is approximately 3-4\% that of the $t$-quark, besides the coupling of the Higgs to bottoms is much smaller than the 400-GeV pseudoscalar one.}

\begin{figure}[t]
\centering
\includegraphics[width=0.45\textwidth]{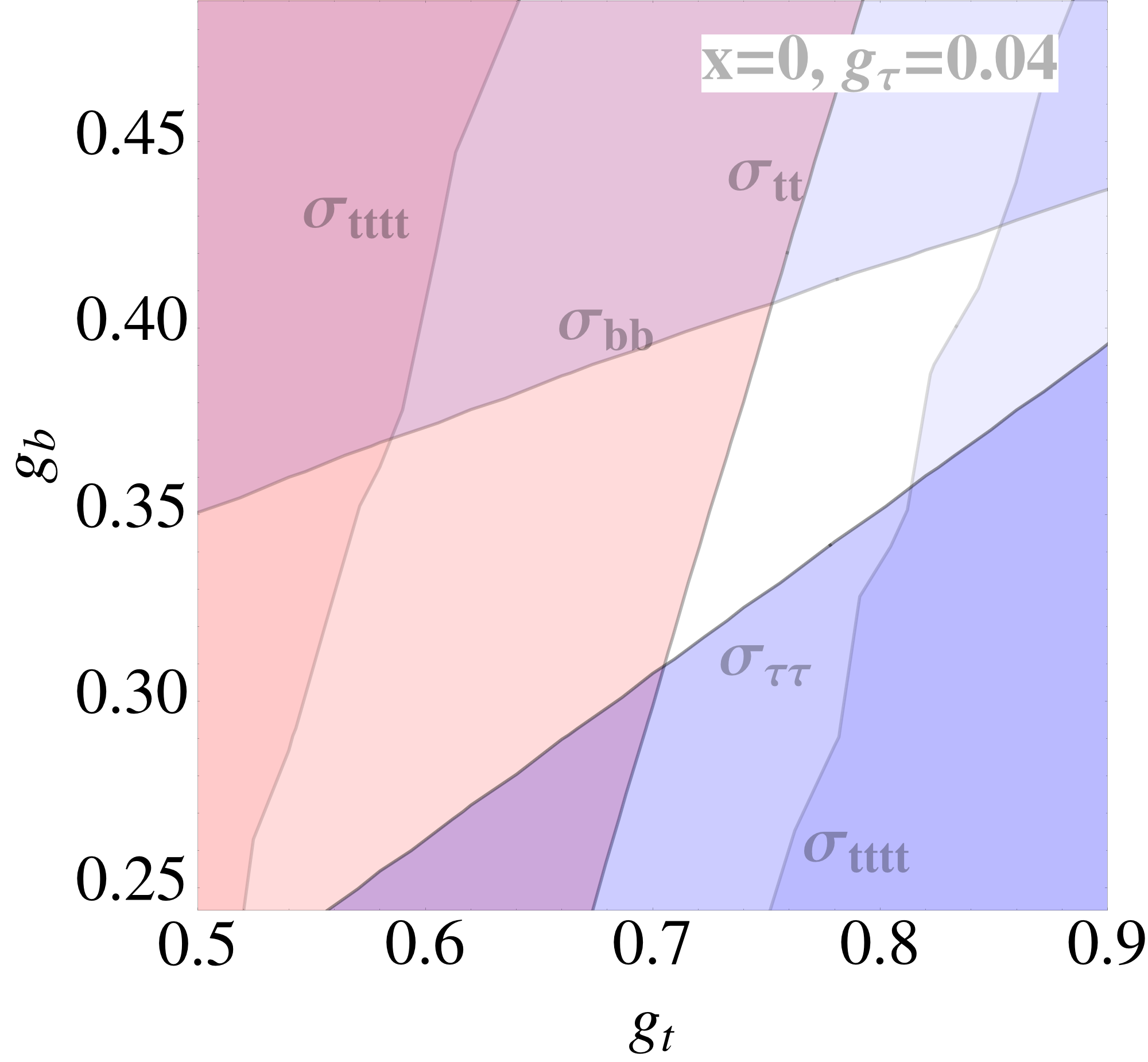}
\includegraphics[width=0.45\textwidth]{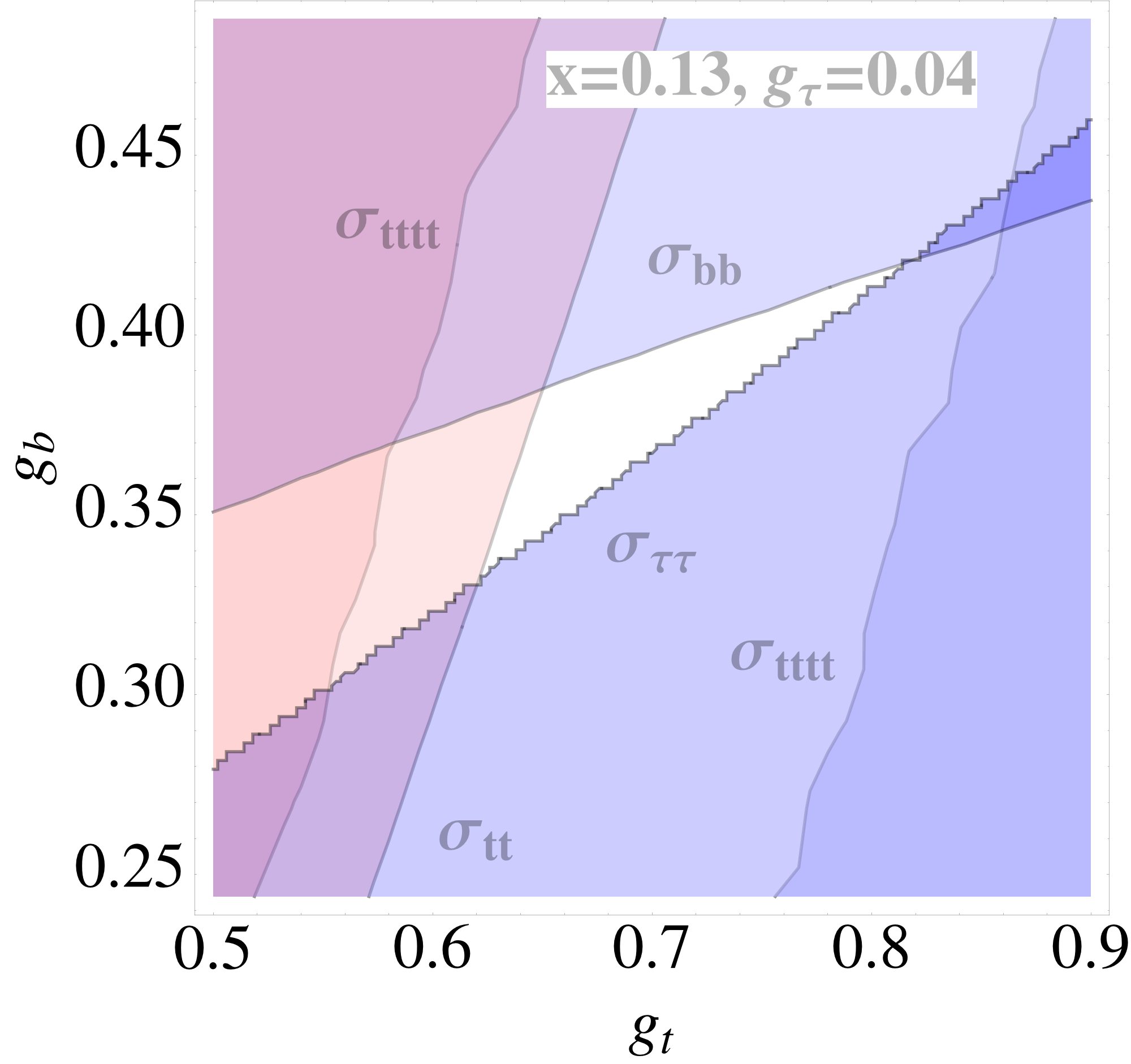}
\caption{Production cross-section predictions of relevant 400-GeV pseudoscalar channels with 1$\sigma$ uncertainties of in the [$g_t$, $g_b$] plane, with $x$ = 0 (left panel) and $x$ = 0.13 (right panel). In both plots $g_\tau$ is fixed to 0.04. Each contour line is labeled with the 1$\sigma$ cross-section value of the corresponding considered channel. Blue and red shaded areas are excluded by data.}
\label{contour-plots_gt-gb}
\end{figure}

\begin{figure}[t]
\centering
\includegraphics[width=0.45\textwidth]{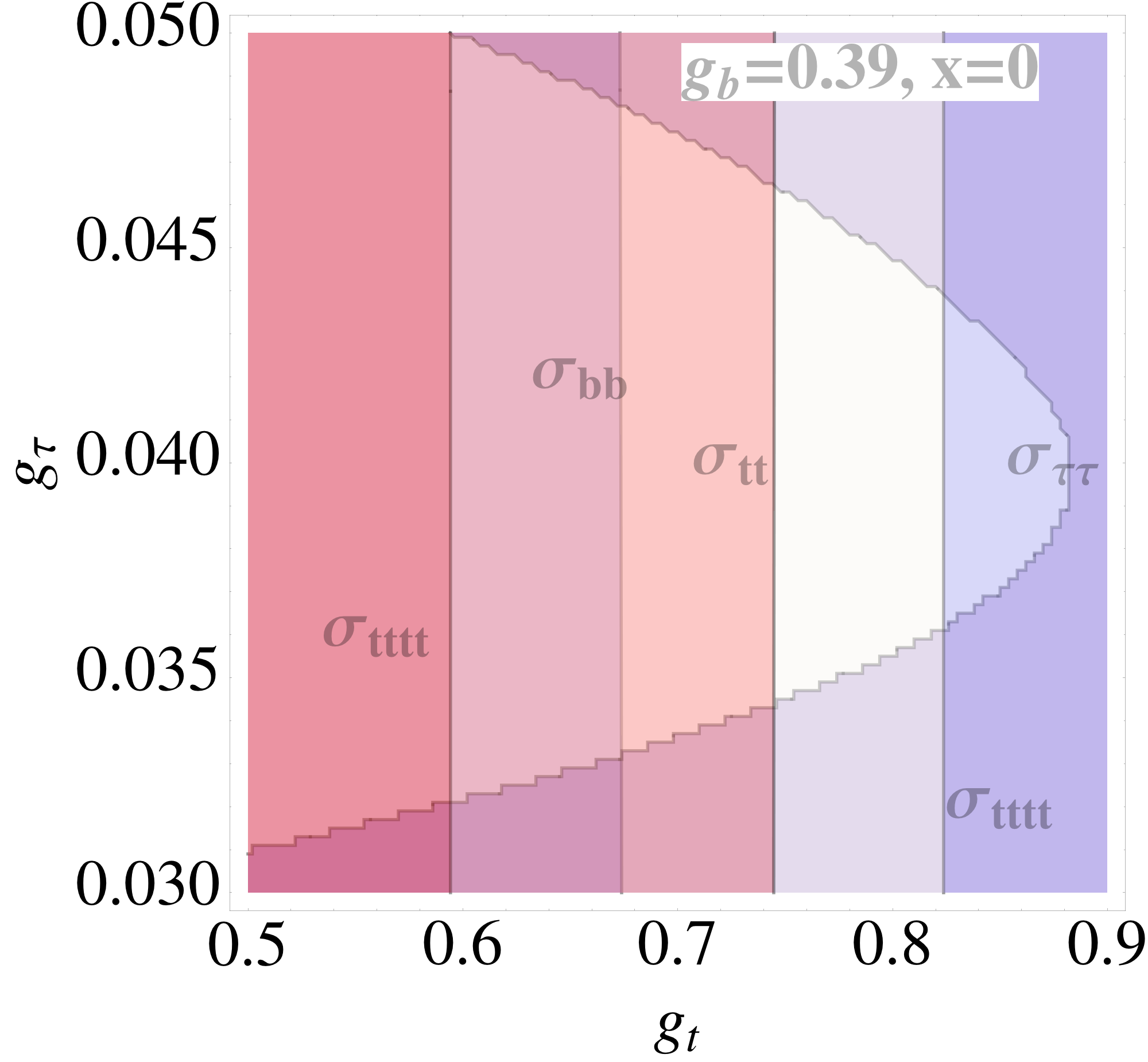}
\includegraphics[width=0.45\textwidth]{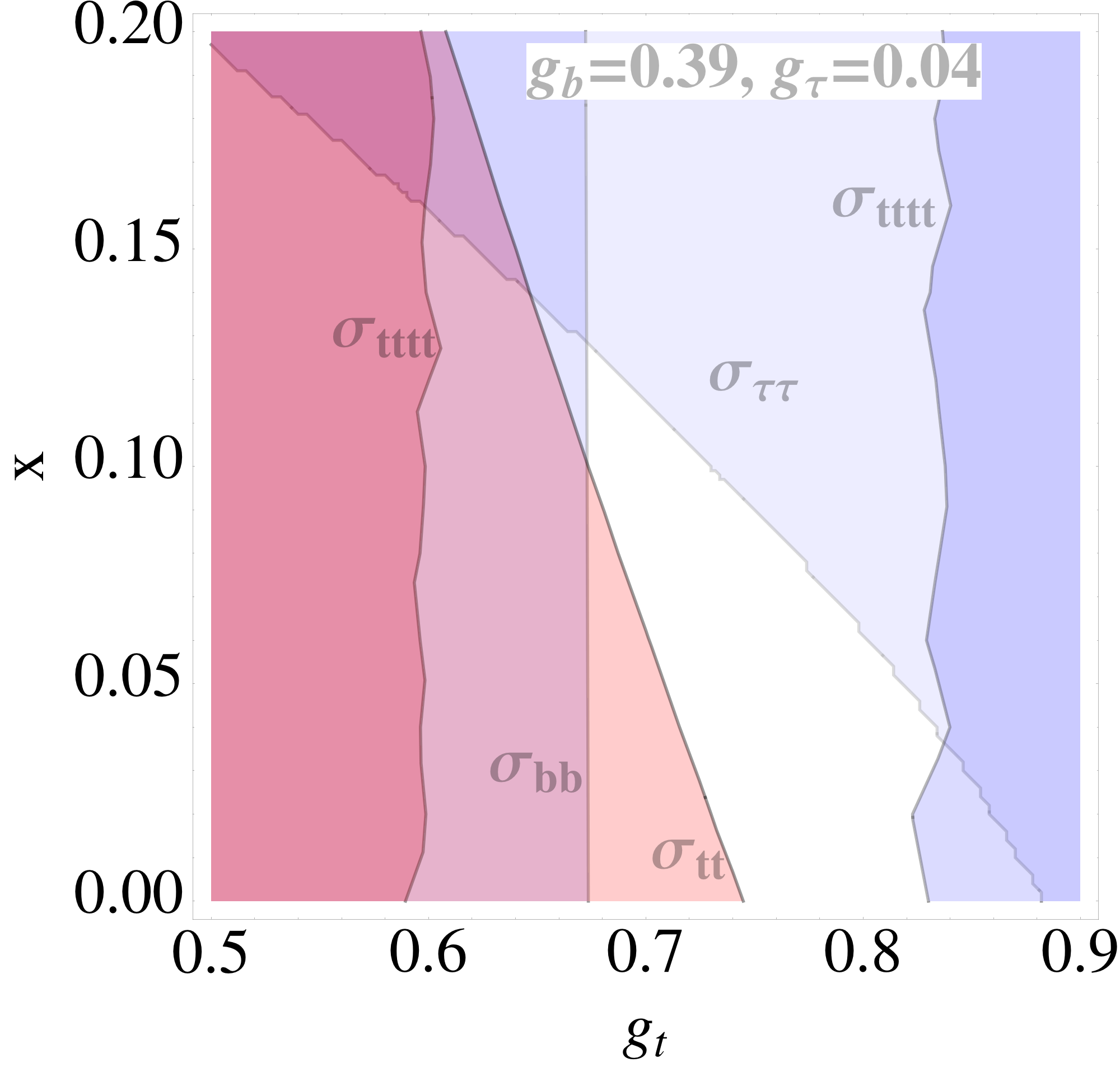}
\caption{Production cross-section predictions of relevant 400-GeV pseudoscalar channels with 1$\sigma$ uncertainties in the [$g_t$, $g_\tau$] (with $x$ = 0 and $g_b$ = 0.39) and [$g_t$, $x$] ($g_{\tau}$ = 0.04 and $g_b$ = 0.39) planes, left and right panels, respectively. Each contour line is labeled with the 1$\sigma$ cross-section value of the corresponding considered channel. Blue and red shaded areas are excluded by data.}
\label{contour-plots_gt-gtau-x}
\end{figure}

In the left panel of Fig.~\ref{contour-plots_gt-gb} we show the values of the cross sections with an uncertainty of 1$\sigma$ of the channels $t \bar t$, $t \bar t t \bar t$, and $\tau^+\tau^-$ in the [$g_t$, $g_b$] plane with $x$ = 0 and $g_\tau$ = 0.04. Each contour line is labeled with the corresponding value at 1$\sigma$, marking the blue (+1$\sigma$) and red (-1$\sigma$) shaded areas that are excluded by data. Therefore, only $g_t$ values ranging from about 0.7 to 0.85 would be allowed by the LHC experimental measurements, provided that $g_b$ is more or less within the range [0.28, 0.4]. In this favored region the width varies 4-6\% with respect to the mass, dominated by the $t \bar t$ decay channel, and the main branching fractions vary among theses values: BR($a$ $\to$ $t \bar t$) $\sim$ 0.6-0.8, BR($a$ $\to$ $b \bar b$) $\sim$ 0.2-0.4, BR($a$ $\to$ $\tau^+ \tau^-$) $\sim$ 0.001, and BR($a$ $\to$ $gg$) $\sim$ 0.035-0.045. These latter values of BR($a$ $\to$ $gg$) hardly change as a functions of the couplings considered and will not be shown again. The right panel of Fig.~\ref{contour-plots_gt-gb} is devoted to same analysis but switching on the coupling $x$, with a value of 0.13. Since the coupling of the pseudoscalar to gluons is now contributing to the $t \bar t$ channel, the constraints on $g_t$ are softer and this coupling can take lower values, close to 0.6. The maximum $g_t$ values are also reduced, up to approximately 0.8. The range of allowed values of $g_b$ is also reduced, from values around from 0.32 to 0.37. In this case the width varies 3.5-6\% and the branching ratios of the dominant channels are BR($a$ $\to$ $t \bar t$) $\sim$ 0.55-0.75, BR($a$ $\to$ $b \bar b$) $\sim$ 0.25-0.45, and BR($a$ $\to$ $\tau^+ \tau^-$) $\sim$ 0.001. The reduction in the white area when comparing the figure on the left with $x=0$ with the right one at $x=0.13$ is provided by the increased constraint in the ditau channel since in fact there is a reduction in the constraint from the ditop channel which would have naively provided a larger white area. However this increment in the ditau constraint can be understood by looking on the right plot of Fig.~\ref{contour-plots_gt-gtau-x} where we see that the ditau ellipse allows larger values of $g_t$ at $x=0$ than at $x=0.13$, which is sensible since both $x$ and $g_t$ contribute to the gluon fusion diagram that enters in the ditau channel. Coming back to Fig.~\ref{contour-plots_gt-gb}, the preferred values for $g_t$ and $g_b$, which we have calculated with our effective model, compatible with the ATLAS and CMS experimental measurements, lie in the white areas and therefore are able to explain the new physics hints while at the same time are consistent with the data at 68\% CL.

Left panel of Fig.~\ref{contour-plots_gt-gtau-x} is dedicated to the pseudoscalar cross-section predictions in the [$g_t$, $g_\tau$] plane, with $x$ = 0 and $g_b$ = 0.39. Contour lines label the values of these cross sections with 1$\sigma$ uncertainty, indicating the blue and red shaded areas not allowed by data. The parameter space region allowed by data is centered at $g_t$ $\sim$ 0.78 and $g_\tau$ $\sim$ 0.040, with $g_t$ varying between 0.74 and 0.82, and $g_\tau$ between 0.033 and 0.047, approximately. In the favored region the width varies 5.5-6\% with respect to the mass, increasing mainly with $g_t$, since the branching fractions to $\tau$-lepton pairs is small. The main branching fractions take the following values: BR($a$ $\to$ $t \bar t$) $\sim$ 0.65-0.7, BR($a$ $\to$ $b \bar b$) $\sim$ 0.3-0.35, and BR($a$ $\to$ $\tau^+ \tau^-$) $\sim$ 0.001-0.002. It is important to remark here that these not-excluded values of $g_t$ are in accordance with the allowed values in the [$g_t$, $g_b$] plane of Fig.~\ref{contour-plots_gt-gb}. Contrary to these contour plots, once the the restrictions from the other channels are accomplished, the $b \bar b$ channel does not impose any restriction on the parameter space of the [$g_t$, $g_\tau$] plane, with the values set as we have indicated for the rest of the couplings, and the most restrictive channels are only $t \bar t$, $\tau^+ \tau^-$, and $t \bar t t \bar t$.

In the right panel of Fig.~\ref{contour-plots_gt-gtau-x} we show the values of the cross sections with an uncertainty of 1$\sigma$ of the channels $t \bar t$, $t \bar t t \bar t$, and $\tau^+\tau^-$ ($gg$ $\to$ $a$ $\to$ $\tau^+ \tau^-$) in the [$g_t$, $x$] plane. Only $g_t$ values ranging from about 0.68 to 0.83 would be allowed by the LHC experimental measurements, provided that $x$ is less than 0.13. In this favored region the width varies 4.5-6\% with respect to the mass, dominated by the $t \bar t$ decay channel, since the branching ratio to gluons is small and the branching ratios of the dominant decay channels are BR($a$ $\to$ $t \bar t$) $\sim$ 0.61-0.68, BR($a$ $\to$ $b \bar b$) $\sim$ 0.31-38, and BR($a$ $\to$ $\tau^+ \tau^-$) $\sim$ 0.001. In this plane, the $b \bar b$ channel comes back into play by excluding a small portion of the parameter space that $t \bar t$, $\tau^+ \tau^-$, and $t \bar t t \bar t$ allowed.

\begin{figure}[t]
\centering
\includegraphics[width=0.45\textwidth]{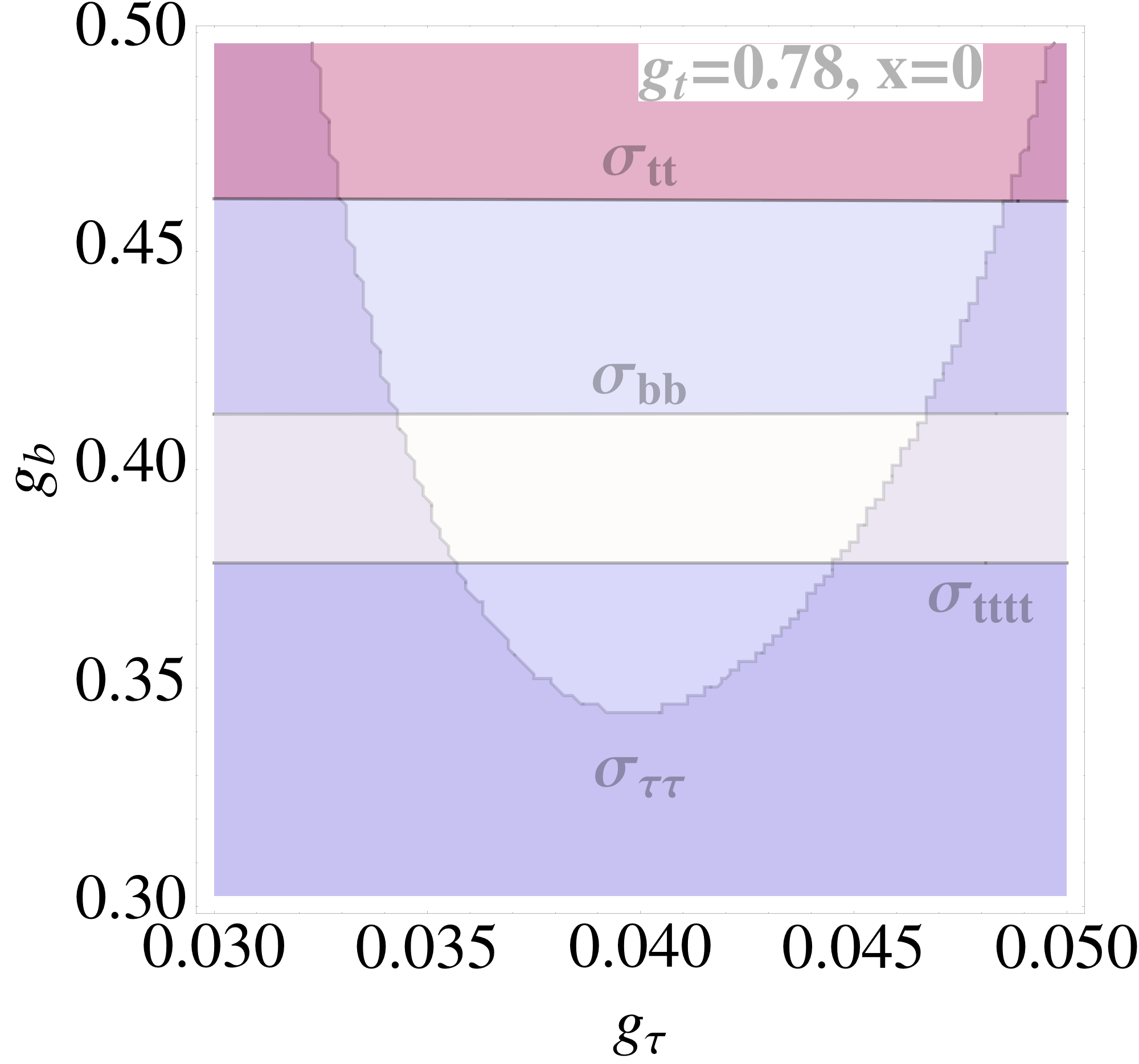}
\includegraphics[width=0.45\textwidth]{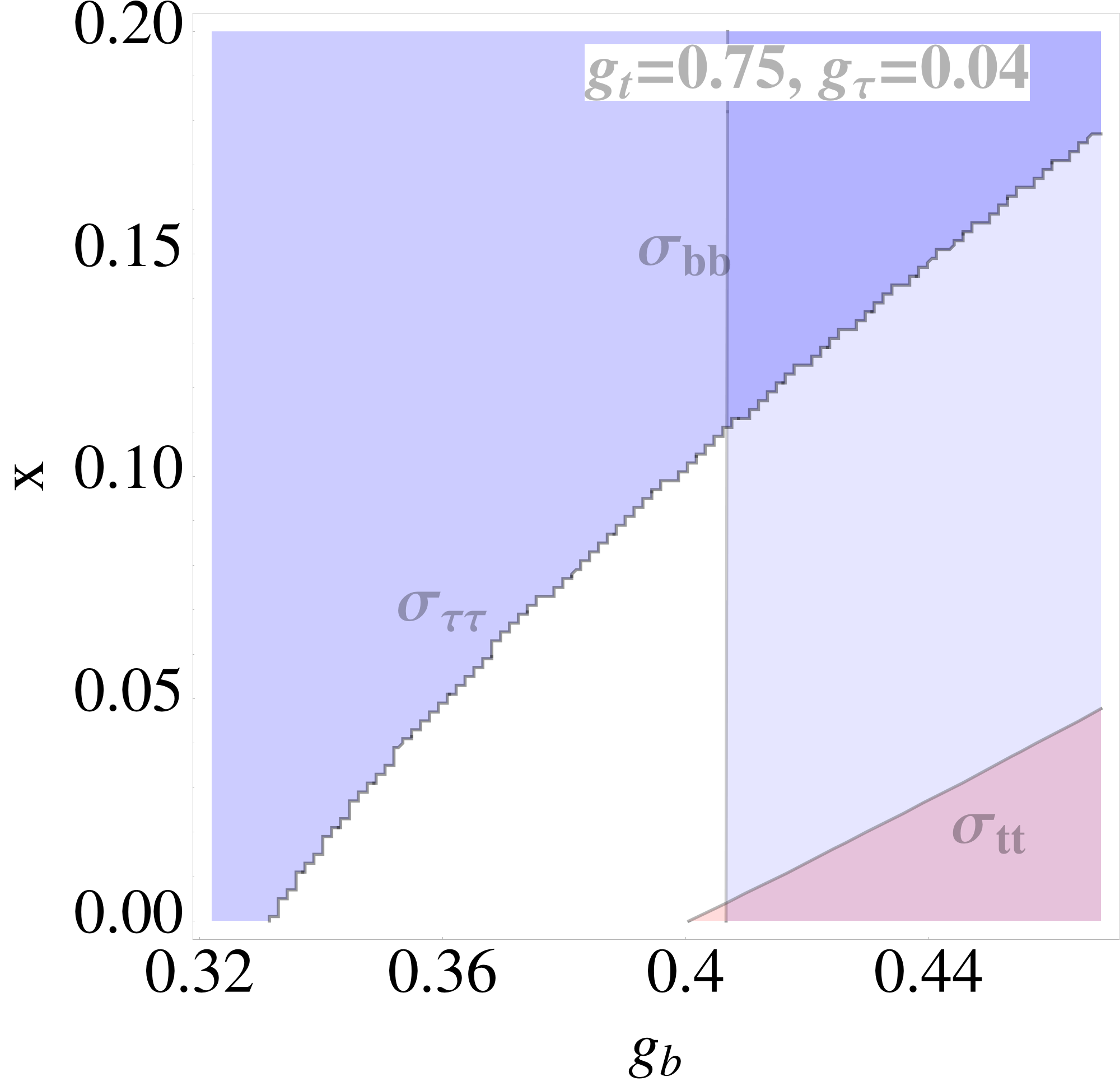}
\caption{Production cross-section predictions of relevant 400-GeV pseudoscalar channels with 1$\sigma$ uncertainties in the [$g_\tau$, $g_b$] (with $g_t =$ 0.78 and $x$ = 0) and [$g_b$, $x$] (with $g_t =$ 0.75 and $g_{\tau}$ = 0.04) planes, left and right, respectively. Each contour line is labeled with the 1$\sigma$ cross-section value of the considered corresponding channel. Blue and red shaded areas are excluded by data.}
\label{contour-plots_gtau-gb-x}
\end{figure}

In the left panel of Fig.~\ref{contour-plots_gtau-gb-x} the 1$\sigma$ cross-section values of the $t \bar t$, $t \bar t t \bar t$, and $b \bar b b \bar b$ channels are displayed in the [$g_\tau$, $g_b$] plane. The allowed values of $g_\tau$ (white area) vary between 0.035 and 0.046, depending on the value of $g_b$, which is constrained to the range [0.37, 0.42]. In the favored region the width varies 5.5-6\% with respect to the mass, increasing mainly with $g_b$, since the branching ratio to $\tau^+ \tau^-$ is small. The branching fraction of the dominant channels are here BR($a$ $\to$ $t \bar t$) $\sim$ 0.65-0.70, BR($a$ $\to$ $b \bar b$) $\sim$ 0.3-0.35, and BR($a$ $\to$ $\tau^+ \tau^-$) $\sim$ 0.001. These values of $g_\tau$ correspond to those allowed in the left panel of Fig.~\ref{contour-plots_gt-gtau-x}, which indicate a good agreement of these data at 68\% CL.

The contour lines of the cross sections of the $t \bar t$, $b \bar b$, and $\tau^+\tau^-$ channels, with 1$\sigma$ of uncertainty, are shown in the right panel of Fig.~\ref{contour-plots_gtau-gb-x} in the [$g_b$, $x$] plane.  Allowed values of $g_b$ vary between 0.33 and 0.4, while $x$ can reach allowed values up to 0.10, depending strongly on $g_b$. In the favored region the width varies 4.5-5.5\% with respect to the mass, increasing mainly with $g_b$, since the branching fraction to gluons is small. The dominant branching ratios are in this case BR($a$ $\to$ $t \bar t$) $\sim$ 0.64-0.72, BR($a$ $\to$ $b \bar b$) $\sim$ 0.28-0.36, and BR($a$ $\to$ $\tau^+ \tau^-$) $\sim$ 0.001. Again, the values of $x$ and $g_b$ are in agreement with the allowed values of Figs.~\ref{contour-plots_gt-gb} and~\ref{contour-plots_gt-gtau-x}.

One significant conclusion is that the most restrictive channels are $t \bar t$, $t \bar t t \bar t$, and $\tau^+ \tau^-$, in which the production of the pseudoscalar and/or its decays are governed by $g_t$, together with the $b \bar b$ channel when the values of $g_b$ are significant and competitive with respect to those of $g_t$. Recall also that in the scenario with $x$ = 0.13, the bounds imposed by $t \bar t t \bar t$ channel are lower than the other channels.

Finally, we find it important to keep in mind that the white areas of the six plots shown in this section are compatible with each other and show the allowed-by-data values for each of the effective couplings of a 400 GeV pseudoscalar that can produce the slight excesses in $t \bar t$ and $\tau^+ \tau^-$ reported by CMS and ATLAS, respectively.

In summary, we find that the ranges of couplings compatible with the excesses in the ditop and ditau channels that at the same time satisfy the constraints from all the other searches mentioned in Section~\ref{exp-searches} are $g_t$ $\in$ [0.68, 0.82], $g_b$ $\in$ [0.3, 0.42], $g_{\tau}$ $\in$ [0.035, 0.047] and $x$ $\in$ [0, 0.13], with small variations depending on the actual value of $x$.

\section{UV completions}
\label{UVcomp}

In this section we describe several models and analyze whether they can reproduce the phenomenology of the previous section. We start with 2HDMs, finding that usually they can not pass the constraints from flavor violating decays. Then we consider models containing a composite pseudoscalar singlet, and study the case where this state is a pNGB.

\subsection{Two Higgs doublet models}
These models contain a neutral pseudoscalar that could reproduce the collider phenomenology. In models of Type-I, II, III and IV the couplings with the SM fermions are proportional to the SM Yukawa coupling, up to the factor $t_\beta$ or its inverse, with $t_\beta$ being defined by the ratio of the vacuum expectation values (vev) of both neutral doublets. In Type-I models all the couplings are given by $\pm t_\beta^{-1}$, in units of the SM Yukawa. In Type-II, up-quark couplings are given by $t_\beta^{-1}$, whereas down-quarks and charged leptons are given by $t_\beta$. In Type-III (also on L 2HDM), quarks have a factor $\pm t_\beta^{-1}$ an charged leptons $t_\beta$. In Type-IV (also in F 2HDM), up-quarks and charged leptons have a factor $\pm t_\beta^{-1}$ and down-quarks a factor $t_\beta$. These structures of couplings cannot accommodate the excesses.

A generalization for the couplings in the Higgs sector of a 2HDM has been considered in Ref.~\cite{Egana-Ugrinovic:2019dqu}  that allow to enhance the interactions with up- or down-type quarks, while suppressing flavor changing neutral currents via flavor alignment. The authors refer to this framework as Spontaneous Flavor Violation (SFV). Particularly interesting for our study is the case of up-type SFV 2HDM, where in the mass basis the up-quark diagonal Yukawa matrix is rescaled by a real parameter $\xi$, whereas each down-type Yukawa is rescaled by an independent factor: $\kappa_i$, namely: $Y_u={\rm diag}(y_u,y_c,y_t)\to\xi Y_u$ and $Y_d={\rm diag}(y_d,y_s,y_b)\to{\rm diag}(\kappa_d,\kappa_s,\kappa_b)$. A similar framework corresponds to the down-type SFV 2HDM, exchanging the role of up- and down-type quarks. In the sector of charged leptons: $Y_\ell={\rm diag}(y_e,y_\mu,y_\tau)\to\xi_\ell Y_\ell$. The new parameters are independent, and can be suitably chosen to reproduce the collider phenomenology we wish to describe.

The problem arises given that this model gets strong constraints from flavor physics. Since the neutral Yukawa couplings are aligned, flavor violation requires the presence of a charged Higgs boson. For large $\kappa_b$ this state gives contributions to $C_7^{bs}$ at 1-loop, inducing $B\to s\gamma$. Following Ref.~\cite{Egana-Ugrinovic:2019dqu}, for the values of $\xi$ and $\kappa_b$ that can fit the results of Section~\ref{num-results}, we obtain a strong lower bound on the mass of the charged Higgs: $m_H\gtrsim 3.4$~TeV. But in 2HDMs $m_H$ and $m_A$ are related by $m_A^2=m_H^2+v^2(\lambda_4-\lambda_5)/2$, with $V\supset \lambda_4(H_2^\dagger H_1)(H_1^\dagger H_2)+\lambda_5(H_1^\dagger H_2)/2$, thus one would have to require $(\lambda_4-\lambda_5)/2\sim{\cal O}(100)$ for $m_A\sim 400$~GeV, loosing perturbative control of the theory.

\subsection{Composite models with a pseudoscalar singlet}

Let us consider now the case of a pseudoscalar singlet state that, given the absence of charged partners, is less constrained than the 2HDM. In particular we find it interesting to consider it as a pseudo NGB associated to a global symmetry of a new sector, since in this case its mass can be naturally smaller than the mass of the other states of the new sector.

We assume the presence of a new strongly interacting sector with a mass gap at an infrared scale of order few TeV, that leads to bound states, to which we will refer as resonances, one of them being a pseudoscalar. We will consider the case where the pseudoscalar is a singlet under the SM gauge symmetry. Moreover, we will take it as a composite NGB of the new sector, generated by the spontaneous breaking of a global symmetry, such that it can be lighter than the rest of the resonances. In this case the shift symmetry of the NGB singlet forbids a potential, making this state massless, but if the symmetry is explicitly broken by the interactions with the SM, a potential is generated at radiative level. The SM gauge interactions can give this explicit breaking of the global symmetry, the interactions with the SM fermions can also do that, the latter being model dependent. One of the most popular examples for the interactions with fermions is partial compositeness, where at a high UV scale $\Lambda_{\rm UV}\gg$TeV linear interactions of the SM fermions and the composite operators are generated: ${\cal L}_{\rm UV}=\lambda\bar\psi^{\rm SM}{\cal O}$.~\footnote{$\Lambda_{\rm UV}$ can be, for example, of order $M_{\rm Pl}$ or $M_{\rm GUT}$.} The shift symmetry can also be broken by anomalous interactions with the SM gauge fields, as in the case of the $\eta'$ of QCD, or by anomalies in the new sector, that are independent from the first ones.~\cite{Gripaios:2007tk,Gripaios:2008ei}

It is interesting to consider that also the Higgs arises as a pNGB~\cite{Kaplan:1983fs,Kaplan:1983sm,Georgi:1984ef,Georgi:1984af,Dugan:1984hq,Contino:2003ve,Agashe:2004rs}, obtaining an explanation for its mass being smaller than the masses of the other resonances, although tuning is still required~\cite{Agashe:2005dk}. Scenarios with the pseudoscalar singlet and the Higgs being composite pNGBs have been considered, for example, in Refs.~\cite{Gripaios:2009pe,Franceschini:2015kwy,Gripaios:2016mmi,Chala:2017sjk}, giving estimates of the potential and couplings with the SM sector, as well as building specific realizations. In the following we will describe some general properties of a pseudoscalar singlet that is a composite pNGB, and in the next subsections we will consider some realizations of these ideas for the hints at 400 GeV. 

In the simplest example the resonances of the strongly interacting sector are determined by one mass scale, namely, the decay constant of the NGBs, $f$, that sets the scales of their self-interactions, and one coupling, $g_*$, that characterizes all the interactions between resonances. These couplings can be thought in analogy with $f_\pi$ and $g_\rho$ for the QCD mesons. We will consider $1\ll g_*\ll4\pi$ and the resonance's masses given by $m_*=g_*f$. There can be departures form this very simplified picture, for example if the NGBs arise from a non-simple group their decay constants can be different: $f_H$ for the Higgs and $f_P$ for the singlet, there can also be different couplings and deviations from the estimate of $m_*$. For simplicity we will assume the one scale and one coupling approach, except where explicitly stated.

For the interactions with fermions we will consider $\Lambda_{\rm UV}\gg m_*$ and approximate scale invariance in a large window of energy, such that the running of the linear coupling $\lambda$ is driven by the anomalous dimension of ${\cal O}$, leading to an exponentially suppressed coupling for an irrelevant operator and to $\lambda\sim g_*$ for a relevant one~\cite{0406257}. At energies of order $m_*$ the linear interactions lead to mixing of the elementary and composite fermions, such that the massless fermions, before EW symmetry breaking, are partially composite, with a degree of compositeness $\epsilon_\psi\sim\lambda_\psi(m_*)/g_*$.
The Yukawa couplings are modulated by $\epsilon$ as: 
\begin{equation}
y_\psi\sim \epsilon_{\psi_L} g_*\epsilon_{\psi_R} \ .
\end{equation}
In the case of three generations of resonances, $\epsilon$ and $g_*$ are squared matrices of dimension three. 

When the breaking of the shift symmetry is dominated by partial compositeness, the mass of the (pseudo) scalar can be estimated as: 
\begin{equation}\label{eq-approx-masses}
m_h^2\sim N_c y_t^2 \frac{g_*^2}{(4\pi)^2} v^2 \ ;
\qquad
m_P^2\sim N_c y_t^2 \frac{g_*^2}{(4\pi)^2} f^2 \ .
\end{equation}
The separation between $v$ and $f$ typically requires a tuning of order $\xi\equiv v^2/f^2$, and allows a splitting between $m_h$ and $m_P$. For $f\simeq 800$~GeV one can obtain $m_P\simeq 400$~GeV, although these estimations are up to factors of ${\cal O}(1)$. \footnote{Contributions to the pseudoscalar mass from anomalies in the new sector can be estimated as
$m_P^2\sim \tilde N_fg_*^2/(4\pi)^2 f^2$,
with $\tilde N_f$ determined by the number of fermions responsible for the anomaly.}

At energies below $m_*$, one gets a theory with the SM fields and the pNGB $H$ and $P$. The $P$ interactions with SM fermions and gauge bosons require at least dimension-5 operators:
\begin{equation}
{\cal L}_{\rm eff}\supset \frac{g_{PAA}}{\Lambda}\ P\ A_{\mu\nu}\tilde A^{\mu\nu}+i \frac{g_{P\psi\psi'}}{\Lambda}P \bar \psi_L H \psi'_R\ ,
\end{equation}
where $A_{\mu\nu}$ stands for a generic SM gauge field strength tensor, $\psi_L$ and $\psi'_R$ are SM fermions and $\Lambda$ is the scale at which these operators are generated, that for simplicity was written to be the same for all the operators. One could also include terms as $\partial_\mu P\bar\psi\gamma^\mu\psi$, however they can be absorbed in the Yukawa and anomalous terms after field redefinitions~\cite{Bellazzini:2015nxw}.

The analysis of Section~\ref{num-results}, as well as the estimates of this section, show that a partially composite third generation of fermions, having sizable couplings with the pNGB pseudoscalar, can generate the proper interactions to account for the sought deviations~\footnote{A model independent analysis using effective operators has been done in~\cite{Banelli:2020iau} in which it was shown that the current LHC searches in 4-tops imply a bound on the composite scale $f \gtrsim 730$ GeV.}. In this scenario it is also possible to generate $PA\tilde A$ interactions by anomalous breaking of the global symmetry of the new sector.
 
\subsubsection{Flavor structure}
Concerning the ditop and ditau deviations, it is enough to couple the pseudoscalar state to the third generation of fermions. However, as we will show below, in this case the pseudoscalar exchange induces FCNC with Wilson coefficients above the bounds allowed by meson physics. Therefore, in the absence of other effects that could ameliorate this situation, one has to consider interactions with the three generations and study this case.

The coupling with the gauge bosons, $PA\tilde A$, breaks the shift symmetry of the NGBs, requiring either insertions of the fermion mixing that explicitly break the symmetry, or an anomalous breaking, generating topological terms, as the Wess-Zumino-Witten term~\cite{Wess:1971yu,Witten:1983tw}~\footnote{See also~\cite{Davighi:2018xwn} for a discussion of topological terms.}. To leading order in $1/f_P$ one gets:
\begin{equation}\label{eq-FFtilde-cp}
{\cal L}_{\rm eff}\supset \frac{P}{16\pi^2 f_P}\left(c_s g_s^2 G\tilde G+c_W g^2 W\tilde W+c_B g'^2 B\tilde B\right) \ .
\end{equation}
The coefficients can be estimated as:
\begin{equation}
\left.c_i\right|_{\rm pc}\sim N_c \frac{y_\psi^2}{g_*^2} \ ,
\qquad 
\left.c_i\right|_{\rm anom}\sim N_f \ .
\end{equation}
In both cases the interaction is generated at loop level, in the first case it is modulated by the Yukawa couplings of the SM fermions, whereas in the second case $c_i$ measures the number of fermions $N_f$ generating the anomaly. 
Rewriting Eq.~(\ref{eq-FFtilde-cp}) in the basis of mass eigenstates, one obtains the interaction with photons, with coefficient $e^2(c_W+c_B)/(16\pi^2f_P)$.

As we will show in the next subsections, it is interesting to write $g_{P\psi\psi}$ as function of $y_\psi$, the matrix of Yukawa coupling of the Higgs, in the following form:
\begin{equation}\label{eq-AB}
\frac{g_{P\psi\psi}}{\Lambda}= \frac{A_\psi y_\psi}{f_P} +\frac{y_\psi B_\psi}{f_P} \ ,
\end{equation}
where $A_\psi$ and $B_\psi$ are $3\times 3$ matrices, determined by the embedding of the Left- and Right-handed fermions, respectively. 
After rotation of the SM fermions to the mass basis, $\psi_{L/R}\to U_{L/R} \psi_{L/R}$, one gets:
\begin{equation}
\frac{g_{P\psi\psi}}{\Lambda}\to U_L^\dagger A_\psi U_L \frac{m_\psi}{f_P} +\frac{m_\psi}{f_P} U_R^\dagger B_\psi U_R\ ,
\end{equation}
where $m_\psi$ is the diagonal fermion mass matrix. As expected, for $A$ and $B$ proportional to the identity the couplings are aligned with the masses, whereas in other cases they are not and lead to flavor transitions.

The analysis of flavor violation by a pNGB pseudoscalar is similar to the analysis of Ref.~\cite{Mrazek:2011iu} for the Higgs sector. If the pseudoscalar and the Higgs arise from the same coset, in the case where there is more than one independent invariant generating the Yukawa interactions, one can expect a misalignment of the of the pseudoscalar couplings. In Section~\ref{so6-so5} we will consider a case where the Right-handed fermions can be embedded in two inequivalent ways, and in Section~\ref{broken-u1} we will consider a case where the coset arises from the product of groups, and the embedding depends on the arbitrary charge under one of these groups. Finding a realization where there is only one independent invariant could lead to alignment of the couplings, relaxing the bounds from flavor.
 
A better determination of the flavor violating couplings requires assumptions about the flavor of the new sector. We will consider the case of flavor anarchy~\cite{Agashe:2004cp}, where the couplings between fermionic and pseudo/scalar resonances are given by matrices with all the coefficients of order $g_*$, such that flavor transitions between resonances have no suppression, and in the conclusions we will comment on other possibilities. It is well known that in this case bounds from flavor changing neutral currents (FCNC) require $m_*\gtrsim 10-20$~TeV~\cite{Agashe:2004cp,Csaki:2008zd}. 

In anarchic partial compositeness the size of the coefficients of the rotation matrices are determined by the degree of compositeness. Ordering the flavors by increasing mixing, leads to $(U_{\psi_L})_{ij}\sim\epsilon_{q^\psi}^i/\epsilon_{q^\psi}^j$ and $(U_{\psi_R})_{ij}\sim\epsilon_{\psi}^i/\epsilon_{\psi}^j$, with $i<j$.

For quarks, assuming that $(U_{u_L})_{ij}\sim(U_{d_L})_{ij}$~\footnote{This assumptions is not necessary, since the condition is $U^\dagger_{u_L}U_{d_L}=V_{CKM}$, it is a simple hypothesis under which we can estimate the mixings. It is also possible to mix the elementary fermions with more than one composite operator, leading to less trivial situations.}, we get:
\begin{align}
&\epsilon_{q}^i\sim (V_{\rm CKM})_{i3}\epsilon_{q}^3 \ , \qquad i<3 \ ,\\
&\epsilon_\psi^i\sim\frac{m_\psi^i}{v}\frac{1}{(V_{\rm CKM})_{i3}g_*\epsilon_{q}^3} \ ,\qquad \psi=u,d \ .
\end{align}
These equations determine the degree of compositeness of the chiral quarks in terms of physical parameters as the CKM angles and masses. The only parameters that are not determined are $\epsilon_{q}^3$ and $g_*$, although the masses give lower bounds for $\epsilon_{q}^3$ when the right mixings are saturated, $\epsilon_q^3\gtrsim 1/g_*$. 

Similar estimates can be made for leptons, although in this case $U_{\rm PMNS}$ depends on the realization of neutrino masses. Assuming that the large mixing angles of the PMNS matrix are generated in the neutrino sector, bounds from flavor violation in the lepton sector are minimized around the Left-Right symmetric limit, $\epsilon_{l}^i\sim\epsilon_e^i$~\cite{Panico:2015jxa}, leading to:
\begin{equation}
\epsilon_{l,e}^i=\sqrt{\frac{m_\ell^i}{g_*v}} \ .
\end{equation}

In this framework the coefficients of the rotation matrices can be estimated as in Table~\ref{table-URij}.
\begin{table}[ht!]
     \centering
     \begin{tabular}{|c|c|c|c|}
       \hline\rule{0mm}{5mm}
	 & $u$ & $d$ & $e$ \\	       
       \hline\rule{0mm}{5mm}
	$(U_R)_{12}$ & $\frac{m_u}{m_c\lambda_C}\sim 0.01$ & $\frac{m_d}{m_s\lambda_C}\sim 0.2$ & $\sqrt{\frac{m_e}{m_\mu}}\sim 0.07$ \\
       \hline\rule{0mm}{5mm}
	$(U_R)_{23}$ & $\frac{m_c}{m_t\lambda_C^2}\sim 0.07$ & $\frac{m_s}{m_b\lambda_C^2}\sim 0.4$ & $\sqrt{\frac{m_\mu}{m_\tau}}\sim 0.2$ \\
       \hline\rule{0mm}{5mm}
	$(U_R)_{13}$ & $\frac{m_u}{m_t\lambda_C^3}\sim 7\times 10^{-4}$ & $\frac{m_d}{m_b\lambda_C^3}\sim 0.08$ & $\sqrt{\frac{m_e}{m_\tau}}\sim 0.02$ \\
       \hline
     \end{tabular}
     \caption{Estimates of coefficients of unitary Right-handed matrices using partial compositeness, taking the masses at the TeV scale.}
     \label{table-URij}
\end{table}

The integration of $P$ leads to FCNC with the four fermion interactions~\cite{Gripaios:2009pe},
\begin{align}
& {\cal L}_{\rm eff}=C_\psi^{ijkl}(\bar\psi^i_L\psi^j_R)(\bar \psi^k_L\psi^l_R) \ ,\label{eq-fcnc1}\\
& C_\psi^{ijkl}=g_{P\psi^i\psi^j}g_{P\psi^k\psi^l}\frac{v^2}{2\Lambda^2}\frac{1}{m_P^2} \ . \label{eq-fcnc2}
\end{align}
The $\Delta F=2$ bounds from meson mixing give constraints on the size of the flavor violating coefficients, $C_{2}^{ij}=C_\psi^{ijij}$.

There can also be dimension-5 operators with covariant derivatives, that can violate flavor, as $P\bar\psi_{L/R} i\not\!\!D \gamma^5\psi_{L/R}$. These operators can be rewritten in terms of $P\bar\psi_L\gamma^5 H\psi_R'$ and its hermitian conjugate after field redefinitions~\cite{Aguilar-Saavedra:2009ygx,Agashe:2009di}.

\subsection{A model with $P$ from SO(6)/SO(5)}
\label{so6-so5}
Refs. \cite{Gripaios:2009pe,Chala:2017sjk} have considered the spontaneous breaking SO(6) to SO(5), that delivers five pNGBs transforming as ${\bf 5}$ SO(5). By using that SO(4)$\sim$SU(2)$_L\times$SU(2)$_R$, under this group ${\bf 5}\sim({\bf 2},{\bf 2})\oplus({\bf 1},{\bf 1})$, leading to a model with custodial symmetry that generates a Higgs and a pseudoscalar singlet. An extra conserved U(1)$_X$ factor must be added to accommodate the hypercharge of the SM fermions, with $Y=T^3_R+X$. 

The NGB unitary matrix is given by:
\begin{equation}
U={\rm exp}(i\sqrt{2}\Pi/f) \ ; \qquad \Pi= h^{\hat a}T^{\hat a} + P T_P \ ;
\end{equation}
where $T^{\hat a}$ and $T_P$ are the broken generators associated to the Higgs and the pseudoscalar.

To determine the interactions of the NGBs with the SM fermions one has to choose the representations of the fermionic composite operators ${\cal O}$ that mix with the elementary fermions. To avoid explicit breaking of the SM gauge symmetry, these SO(6) representations, when decomposed under the SM gauge symmetry, must contain multiplets transforming as the SM fermions, or in other words: the EW representations of the SM fermions are embedded into representations of SO(6)$\times$U(1)$_X$. These embeddings are not unique, and the phenomenology of $P$, in many aspects, depends strongly on this choice. We will consider the following embeddings:
\begin{equation}\label{eq-embeddings1}
{\cal O}_{q^u}, {\cal O}_u \sim {\bf 6}_{2/3} \ ,\qquad {\cal O}_{q^d},{\cal O}_d\sim {\bf 6}_{-1/3} \ , 
\qquad
\psi_l, \psi_e \sim {\bf 6}_{-1} \ .
\end{equation}
The presence of two embeddings for $q$: $q^u$ and $q^d$, means that it interacts linearly with two different operators: ${\cal L}\supset \bar q(\lambda_{q^u}{\cal O}_{q^u}+\lambda_{q^d}{\cal O}_{q^d})$.
{\bf6} is the smallest dimensional representation that allows to protect the $Zb_L\bar b_L$ coupling, that demands embedding $q$ in a $({\bf 2},{\bf 2})_{2/3}$ of SO(4)$\sim$U(1)$_X$~\cite{0605341}. $u$ is embedded in the same representation, but since  there are two SO(4) singlets ${\bf 6}\sim{\bf5}\oplus{\bf1}\sim({\bf 2},{\bf 2})\oplus({\bf 1},{\bf 1})\oplus({\bf 1},{\bf 1})$, there is a free parameter $c_u\equiv\cos\theta_u$ that determines the projection over the SO(5) singlet, whereas $s_u\equiv\sin\theta_u$ determines its projection over the {\bf5}, in fact there is one $\theta_u$ for each generation. Since ${\bf 6}_{2/3}$ does not contain a -1/3 singlet, the down sector is embedded in a ${\bf 6}_{-1/3}$, with a new parameter $\theta_d$, similar to the up sector. Finally the leptons are embedded in a ${\bf 6}_{-1}$, with a $\theta_e$ for the lepton singlet.

Since the SM fermions do not fill the embeddings, the linear interactions that lead to partial compositeness explicitly break the global symmetry and generate a potential for the NGBs, which coefficients can be estimated as we did for the quadratic ones in Eq.~(\ref{eq-approx-masses})~\cite{Chala:2017sjk}.

We assume that at low energies, of order few to ten TeV, the strong dynamics generates resonances 
that mix with the elementary fermions. 
Integrating them one obtains to leading order:
\begin{equation}\label{eq-lyso6}
{\cal L}_{\rm eff}\supset\sum_{\psi=u,d,\ell}\bar \psi_L \frac{m_\psi}{v}\left(h+i\frac{P}{f}\cot_\psi\right)\psi_R + {\rm h.c.} \ ,
\end{equation}
where generation indices are understood, such that $\cot_\psi$ is a matrix in the generation space: 
\begin{equation}
\cot_\psi={\rm Diag}(\cot\theta_{\psi}^1,\cot\theta_{\psi}^2,\cot\theta_{\psi}^3) \ ,\qquad \cot\theta_{\psi}=c_\psi/s_\psi \ . 
\end{equation}
Comparing with Eq.~(\ref{eq-AB}), in this model we obtain $A=I$ and $B=\cot_\psi$, a diagonal matrix. 

Rotating the chiral fermions to the mass basis, the diagonal coefficients of the second term of Eq.~(\ref{eq-lyso6}) lead to:
\begin{equation}\label{eq-yso6}
g_{P\psi^i\psi^i}\frac{v}{\Lambda}=\frac{m_{\psi^i}}{f}\sum_j|U_{\psi_R}|_{ji}^2\cot\theta_\psi^j\simeq\frac{m_{\psi^i}}{f}\cot\theta_\psi^i \ ,
\end{equation}
where in the right-hand-side we have made the assumption that the term with $j=i$ dominates, that will be the case when the mixing angles are small and the ratio $\cot\theta_\psi^j/\cot\theta_\psi^i$ does not compensate the smallness of those mixings.

Although $\cot_\psi$ is diagonal, if its elements are not universal, it does not commute with $U_{\psi_R}$ and generates flavor violating interactions $y_{P\psi^i\psi^j}$. For $i\neq j$:
\begin{equation}
(U_{\psi_R}^\dagger\cot_\psi U_{\psi_R})_{ij}=(\cot_{\psi_2}-\cot_{\psi_1})(U_{\psi_R}^*)_{2i}(U_{\psi_R})_{2j}+(\cot_{\psi_3}-\cot_{\psi_1})(U_{\psi_R}^*)_{3i}(U_{\psi_R})_{3j} \ .
\end{equation}

Since SO(6)$\sim$SU(4), it contains anomalous representations. Fermions of the composite sector transforming with these representations contribute to the anomalous interactions with the fields of SO(6):
\begin{equation}
c_s|_{\rm anom}=0 \ ,\qquad c_W|_{\rm anom}=-c_B|_{\rm anom}=n \ ,
\end{equation}
where $n$ is an integer number. Unfortunately, in this model there are no anomalous contributions to the interactions with gluons, that could increase $\sigma(gg\to P)$ with small impact on $\sigma_{4t}$, neither with the photons, that could boost a signal in the clean $\gamma\gamma$ decay channel.

\subsubsection{Matching}
Making use of the previous results we can obtain the couplings of Eq.~(\ref{eq-Lint}) in the SO(6)/SO(5) model. From Eqs.~(\ref{eq-lyso6}) and (\ref{eq-yso6}) we get:
\begin{equation}
g_{\psi^i}
\simeq \frac{m_\psi^i}{f} \cot\theta_\psi^i \ , 
\qquad
g_{gg} 
\simeq 4.2\frac{g_s^2}{16\pi^2f} \cot\theta_t\ .
\end{equation}
Using the benchmark point with $x=0$, described in Section~\ref{num-results}, we get for the third generation:
\begin{align}
& \cot\theta_t \sim 4 \frac{f}{\rm TeV} \ , 
\qquad \cot\theta_b \sim 145 \frac{f}{\rm TeV} \ , 
\qquad \cot\theta_\tau \sim 23 \frac{f}{\rm TeV} \ .
\end{align}

Notice that although $\cot\theta\sim 100$, it does not lead to large couplings with the pseudoscalar, since these couplings are proportional to the quark masses. In the case of flavor universal $\cot\theta$ one gets: $g_{d}\sim 2\times 10^{-4}$, $g_{s}\sim 2.5\times 10^{-3}$ and $g_{b}\sim 0.3$.

Thus by choosing these values for $\cot\theta$, the SO(6)/SO(5) model is able to reproduce the phenomenology described in the previous sections, in the case without anomalous interactions from the new sector.

\subsubsection{Flavor transitions}
Let us study now FCNC induced by $P$-exchange. The Wilson coefficient of Eq.~(\ref{eq-fcnc2}) is given by
\begin{equation}
C_\psi^{ijkl}=\frac{1}{m_P^2}\frac{m_\psi^i}{f}\frac{m_\psi^k}{f}(U_{\psi_R}^\dagger\cot_\psi U_{\psi_R})_{ij}(U_{\psi_R}^\dagger\cot_\psi U_{\psi_R})_{kl} \ .
\end{equation}
Using the estimates of Table~\ref{table-URij} and Ref.~\cite{0707.0636} we can obtain bounds on
\begin{equation}
c_\psi^{(ij)} = \cot_\psi^i-\cot_\psi^j \ .
\end{equation}
Taking $f=1$~TeV and the Wilson coefficient as $C=(c/{\rm TeV})^2$, we get the predictions and bounds of Table~\ref{table-bounds-q}. 
\begin{table}[ht!]
     \centering
     \begin{tabular}{|c|c|c|c|}
       \hline\rule{0mm}{5mm}
&\multicolumn{1}{c}{Model prediction}&\multicolumn{2}{|c|}{Bounds} \\
	 & $c$ & $\sqrt{{\rm Re}(c^2)}$ & $\sqrt{{\rm Im}(c^2)}$ \\
       \hline\rule{0mm}{7mm}
	$K^0$ & $2\times 10^{-5}c_d^{(21)}+3\times 10^{-6}c_d^{(31)}$ 
& $1.4\times 10^{-4}$ & $1\times 10^{-5}$ \\
       \hline\rule{0mm}{7mm}
	$D^0$ & $10^{-5}c_u^{(21)}+5\times 10^{-8}c_u^{(31)}$ 
& $4.4\times 10^{-4}$ & $4.4\times 10^{-4}$ \\
       \hline\rule{0mm}{7mm}
	$B_d^0$ & $4\times 10^{-4}(c_d^{(21)}+c_d^{(31)})$ 
& $8\times 10^{-4}$ & $8\times 10^{-4}$ \\
       \hline\rule{0mm}{7mm}
	$B_s^0$ & $2\times 10^{-3}(c_d^{(21)}+c_d^{(31)})$ 
& $8\times 10^{-3}$ & $8\times 10^{-3}$ \\
       \hline
     \end{tabular}
     \caption{Wilson coefficients for meson mixing. Defining the Wilson coefficient as $C=(c/{\rm TeV})^2$ an taking $f=1$~TeV, the second column shows the estimates for $c$ in the model, the third and fourth columns show the bounds on the real and imaginary parts of these coefficients~\cite{0707.0636}.}
     \label{table-bounds-q}
\end{table}

From Table~\ref{table-bounds-q} and assuming real Wilson coefficients, we obtain that the values $c_d^{(21)}\sim 5$ and  $c_d^{(31)}\sim 34$ saturate the bound from $K$-system, whereas $B_d$ requires $c_d^{(21)}+c_d^{(31)}\lesssim 2$ and $B_s$: $c_d^{(21)}+c_d^{(31)}\lesssim 4$ for saturation. Since $\cot\theta_b\sim 100$, constraints from $B$-meson mixing require a cancellation in $c_d^{(ij)}$ of order few percent, with one more order of magnitude for complex coefficients with arbitrary phases. 
 
Bounds from $D$ mesons allow $c_u^{(21)}\sim 44$, while constraints on $c_u^{(31)}$ are very soft. Since $\cot\theta_t\sim 5$, no tuning is needed to satisfy these bounds.

For leptons, there are strong bounds from $\mu\to 3e$ and $\tau\to \ell\ell'\ell''$, with $\ell$ being muons and electrons. However, since the couplings involved are suppressed by the initial and final lepton masses, for $\cot\theta_e^i\lesssim{\cal O}(30)$ the predicted flavor violating BRs are several orders of magnitude below the bounds.

Summarizing, departures from universality must be lower than few percent (per mil) for the down sector in the case of real (complex) coefficients, demanding either tuning or flavor symmetries and departure from anarchy, whereas they can be of order one for the up sector and leptons.

\subsection{A model with $P$ from a broken U(1)}
\label{broken-u1}
The coset SO(5)$\times$U(1)$_P\times$U(1)$_X$/SO(4)$\times$U(1)$_X$ also delivers the Higgs and the pseudoscalar as NGBs from a strongly interacting sector.~\footnote{A conserved SU(3) factor accounting for color is understood.} The Higgs emerges from the same coset as in the well known minimal composite Higgs model, namely, SO(5)/SO(4), whereas $P$ emerges from the spontaneous breaking of U(1)$_P$. The extra U(1) factor is not spontaneously broken and it is required to accommodate the hypercharge of the SM fermions, with $Y=T^3_R+X$. Since the NGBs arise from different factors of the symmetry group, $H$ and $P$ can have different decay constants. We assume that the same dynamics is responsible for the spontaneous breaking of both factors and take them of the same order.

The NGB unitary matrices are given by:
\begin{equation}
U_H={\rm exp}(i\sqrt{2}h^{\hat a}T^{\hat a}/f_H) \ ; \qquad U_P= {\rm exp}(i\sqrt{2}PQ_P/f_P) \ ;
\end{equation}
where $T^{\hat a}$ and $Q_P$ are the broken generators of SO(5) and U(1)$_P$. For a state with well defined charge $Q_P$, $U_P$ is just a phase.

In order to determine the interactions of the NGBs with the SM fermions, one has to choose an embedding into representations of SO(5)$\times$U(1)$_P\times$U(1)$_X$. A simple and realistic case can be obtained by considering the following:
\begin{equation}\label{eq-embeddings}
{\cal O}_{q,u}\sim {\bf 5}_{2/3,p_{q,u}} \ ,\qquad {\cal O}_d\sim {\bf 10}_{2/3,p_d}, 
\qquad
{\cal O}_l\sim {\bf 5}_{0,p_l} \ ,\qquad {\cal O}_e\sim {\bf 10}_{0,p_e} \ . 
\end{equation}
The first and second subscripts in Eq.~(\ref{eq-embeddings}) are the U(1)$_X$ and U(1)$_P$ charges, respectively. The values of $Q_P$ are {\it a priori} arbitrary. 

Again the linear interactions of partial compositeness explicitly break the global symmetry and generate a potential for the NGBs. Ref.~\cite{Chala:2017sjk} has shown that, to generate a potential for $P$, at least one SM fermion must interact with two operators of the new sector with different $p$-charges. Following that reference we introduce two embeddings for $u_R$, with different charges under U(1)$_P$: $p_u^1$ and $p_u^2$.

At energies below the TeV, where the massive resonances can be integrated, ones obtains Yukawa interactions:
\begin{align}\label{eq-yukawa}
{\cal L}_{\rm eff}\supset &\bar u_L  u_R\left[\frac{h}{v}m_u +i\sqrt{2}\frac{P}{f_P}\left(p_qm_u -m_u p_u\right)\right]
+ \bar d_L d_R\left[\frac{h}{v}m_d +i\sqrt{2}\frac{P}{f_P}(p_qm_d - m_dp_d)\right]
\nonumber\\
&+ \bar e_L e_R\left[\frac{h}{v}m_e +i\sqrt{2}\frac{P}{f_P}(p_lm_e-m_ep_e)\right]
\end{align}
with $p_u=(p_{u1}+p_{u2})/2$. In the case of one generation only, to leading order the $P$ couplings are proportional to the Higgs ones, being determined by the $Q_P$ charges and $f_P$. In the case of three generations $p_\psi$ are diagonal matrices, thus the alignment with the Higgs coupling depends on whether $Q_P$ is universal for the three generations or not. Comparing with Eq.~(\ref{eq-AB}), in this model we obtain $A=\sqrt{2}p_{\psi_L}$ and $B=-\sqrt{2}p_{\psi_R}$. 

If the diagonal terms dominate the sums, the flavor conserving coupling can be approximated by:
\begin{equation}\label{eq-yso5u1}
g_{P\psi^i\psi^i}\frac{v}{\Lambda}=\sqrt{2}\frac{m_{\psi^i}}{f}\sum_j(|U_{\psi_L}|_{ji}^2p_{\psi_L}^{(j)}-|U_{\psi_R}|_{ji}^2p_{\psi_R}^{(j)})\simeq\sqrt{2}\frac{m_{\psi^i}}{f}(p_{\psi_L}^{(i)}-p_{\psi_R}^{(i)}) \ .
\end{equation}
The flavor violating couplings have now contributions from Left- and Right-handed unitary matrices, for $i\neq j$:
\begin{align}\label{eq-gijso5u1}
g_{P\psi^i\psi^j}\frac{v}{\Lambda}=
&[(p_{\psi_L}^2-p_{\psi_L}^1)(U_{\psi_L}^*)_{2i}(U_{\psi_L})_{2j}+(p_{\psi_L}^3-p_{\psi_L}^1)(U_{\psi_L}^*)_{3i}(U_{\psi_L})_{3j}]\frac{m_\psi^j}{f_P} \nonumber 
\\ 
&-[(p_{\psi_R}^2-p_{\psi_R}^1)(U_{\psi_R}^*)_{2i}(U_{\psi_R})_{2j}+(p_{\psi_R}^3-p_{\psi_R}^1)(U_{\psi_R}^*)_{3i}(U_{\psi_R})_{3j}]\frac{m_\psi^i}{f_P} \ .
\end{align}

The triangle anomaly with one U(1)$_P$ and two SU(3)$_c$ or SO(5) generators gives anomalous couplings~\cite{Franceschini:2015kwy}, for a multiplet of composite fermions $f$:
\begin{equation}\label{eq-c-matching-cp}
c_s|_{\rm anom}=\sum_f p_f d_2^{(f)} I_3^{(f)} \ , \quad 
c_W|_{\rm anom}=\sum_f p_f d_3^{(f)} I_2^{(f)} \ , \quad 
c_B|_{\rm anom}=\sum_f p_f d_2^{(f)} d_3^{(f)} (Y^{(f)})^2 \ ,
\end{equation}
where $d_2$ and $d_3$ ($I_2$ and $I_3$) are the dimensions (indices) of the representations under SU(2)$_L$ and SU(3)$_c$, respectively. In the present example, for each generation of fermionic resonances transforming as in Eq.~(\ref{eq-embeddings}), taking $p_f=1$ for all the fermions, one obtains: $c_s\simeq 12$, $c_W\simeq 22$ and $c_B\simeq 14$. 

\subsubsection{Matching}
To reproduce the phenomenology, we match the couplings of Eq.~(\ref{eq-Lint}) in the SO(5)$\times$U(1)/SO(4) model. From Eqs.~(\ref{eq-yukawa}) and~(\ref{eq-yso5u1}) we obtain:
\begin{equation}
g_{t}
\simeq \sqrt{2}\frac{m_{t}}{f_P} (p_q^{(3)}-p_u^{(3)}) \ ,
\qquad
g_{b}
\simeq \sqrt{2}\frac{m_{b}}{f_P} (p_q^{(3)}-p_d^{(3)}) \ ,
\qquad
g_{\tau}
\simeq \sqrt{2}\frac{m_{\tau}}{f_P} (p_l^{(3)}-p_e^{(3)}) \ .
\end{equation}
Taking into account Eq. (\ref{eq-FFtilde-cp}), we obtain for the anomalous coupling to gluons,
\begin{align}
g_{gg} \simeq \frac{g_s^2}{16\pi^2f_P} \left[4.2\sqrt{2}(p_q^{(3)}-p_u^{(3)}) + 4 c_s|_{\rm anom}\right] \ ,
\end{align}
with $c_s|_{\rm anom}$ given in Eq.~(\ref{eq-c-matching-cp}). Similar expressions can be derived for EW gauge bosons.

For the benchmark points we get:
\begin{align}\label{eq-pvalues}
 p_q^{(3)}-p_u^{(3)} \simeq 2.6 \frac{f_P}{\rm TeV} \ , \qquad 
 p_q^{(3)}-p_d^{(3)} \sim 10^2 \frac{f_P}{\rm TeV} \ , \qquad 
 p_l^{(3)}-p_e^{(3)} \sim 20 \frac{f_P}{\rm TeV} \ .
\end{align}

The bottom-quark requires charges ${\cal O}(100)$. The global current associated to the symmetry U(1)$_P$ is expected to create spin one composite states, that can couple to the composite fermions with coupling $\tilde g_*$ and charge $Q_P$. A well defined perturbative description of the theory of resonances requires $\tilde g_* Q_P\ll 4\pi$, thus for charges ${\cal O}(100)$ one has to demand $\tilde g_* \ll 0.1$. However, since the mass of the spin-one state is estimated as $\sim g_* f_P$, for such a small U(1)$_P$ coupling one would expect a very light resonance. This is in fact expected given that for such a large charge the UV-running would imply a Landau pole at low energies, signaled by the presence of low mass resonances. We consider this to be a more serious problem than the flavor constraints discussed in the next section, that depend on several assumptions in the UV. Therefore, this model is in strong tension with the large coupling $g_b$, that is required at 68\% CL by the $\tau$-channel with pseudoscalar production by bottom fusion. However this strong tension can be alleviated by considering the 95\% CL limits, that contain a region with $g_b=0$ and $g_\tau\gtrsim 0.023$, requiring $(p_l^{(3)}-p_e^{(3)})\sim 10$.

For the second benchmark point, where an anomalous contribution to the gluon coupling is included, taking $x\sim 0.13\pm 0.05$ we obtain:
\begin{equation}
\sum_f p_f d_2^{(f)} \simeq (1.4-2.7) \frac{f_P}{\rm TeV} \ .
\end{equation}

\subsubsection{Flavor}
Using the couplings of Eq.~(\ref{eq-gijso5u1}) in~(\ref{eq-fcnc2}) one obtains the Wilson coefficients induced by $P$-exchange in this model.
The bounds on deviations from universality of $p_{\psi_R}^{(i)}$ are as in Table~\ref{table-bounds-q}, changing $c_\psi^{(ij)}$ by $\Delta p_{\psi_R}^{(ij)}\equiv(p_{\psi_R}^{(j)}-p_{\psi_R}^{(i)})$. Thus the constraints on $c_\psi^{(ij)}$ apply also to $\Delta p_{\psi_R}^{(ij)}$. For leptons, in the Left-Right symmetric limit, constraints on $\Delta p_{l}^{(ij)}$ are are similar to those on $\Delta p_{e}^{(ij)}$. 

In this model there are as well bounds on $\Delta p_{\psi_L}^{(ij)}$, that can be obtained by a similar calculation, taking into account that the Left-handed angles are of CKM size. Assuming real Wilson coefficients, for the Kaon system we get: $\Delta p_q^{(21)}\lesssim 7$ and  $\Delta p_q^{(31)}\lesssim 3\times10^3$, for $B_d$: $\Delta p_q^{(21)},\Delta p_q^{(31)}\lesssim 17$ and for $B_s$: $\Delta p_q^{(21)},\Delta p_q^{(31)}\lesssim 34$. Since the estimate of Eq.~(\ref{eq-pvalues}) gives $(p_q^{(3)}-p_d^{(3)})\gtrsim 10^2$, the constraints from meson mixing require cancellations at few percent level, whereas for complex coefficients the tuning is at per mil level. 
For $D$ mesons $\Delta p_q^{(21)}\lesssim 2$, while constraints on $\Delta p_q^{(31)}$ are very soft. 


\section{Conclusions}
\label{conclu}

Several experimental results from the LHC could be pointing to a new pseudoscalar state at 400 GeV, mostly coupled to the third generation of fermions. Results from CMS in the $t\bar t$ final state as well as the results from ATLAS in the $\tau^+\tau^-$ channel, favor a pseudoscalar neutral particle that is produced via gluon fusion or in association with bottom-quarks. We have analyzed from a phenomenological perspective, initially in a bottom-up approach, the couplings that such state should have with those fermions and gauge bosons to be able to reproduce the different excesses, while at the same time keeping below the bounds the related channels that show no deviations with respect to the SM. Scanning over the parameter space of the most general CP-invariant interaction Lagrangian linear in the pseudoscalar state $a$ that can be written up to dimension-5 operators, we have found regions in which the new physics hints, as well as all experimental constraints, can be satisfied and in which the pseudoscalar coupling to gluons can be induced by the SM top quark, that gives by far the dominant contribution, but there can also be room for contributions from heavy new states. We found that the couplings of the pseudoscalar to top quarks and $\tau$ leptons are of the same order of magnitude as the Higgs ones, while the coupling to bottom-quarks is required to be $\sim 20-30$ times larger than the one for the Higgs.  

Though our low-energy phenomenological effective model satisfies all current experimental bounds and provides an explanation for the hints in ditop and ditau final states for the regions of parameter space considered, one may wonder in which of these and/or other channels one could expect to find a signal for the presence of the pseudoscalar in future measurements.  In this aspect, one of the channels in which one would expect to be able to probe the previous scenario in the future is in the production of 4-tops in which there is currently an excess that is around twice the SM one. Furthermore, given the hint of an excess in the $\tau^+\tau^-$ channel initiated by $b$-quarks, one could also expect the pseudoscalar to provide an excess in the $b\bar b$ final states in future measurements. On the other hand diboson channels are more model dependent due to their possible UV contributions, thus there are no robust predictions for diboson final states, though one would expect any possible signal to show first in the diphoton channel due to its cleanness.

We have also considered a set of gauge invariant models, analyzing their capability to reproduce the collider phenomenology related with the hints at 400 GeV. We have found that, although usual 2HDMs of Type I-IV cannot reproduce the pseudoscalar couplings required by the phenomenology, more sophisticated models as SFV 2HDM could in principle do it~\cite{Egana-Ugrinovic:2019dqu}. However the contributions of this model to flavor changing  processes, induced at radiative level by exchange of a charged Higgs, require the masses of these states to be above $\sim 3$~TeV, demanding quartic couplings above the perturbative regime to split the pseudo scalar mass $m_A$ from the CP-even heavy Higgs mass $m_H$. On a different direction, models with a new strongly interacting sector leading to a light pseudoscalar singlet could also potentially reproduce the LHC phenomenology we wish to address. We have analyzed two specific realizations that have already been considered in the literature, although in a different context. One of the models contains a composite state that is a pseudo Nambu Goldstone boson arising from an spontaneous breaking of SO(6)/SO(5), which Yukawa couplings are aligned with the Higgs ones. In this model SO(6) anomalies can generate contributions to the couplings with EW massive gauge bosons, but not with the gluons or photons. Under the assumption of flavor anarchy, bounds from mixing in Kaon and $B$-meson systems require tuning to obtain an approximate universal factor in the down sector, another interesting possibility that requires a dedicated analysis would be the introduction of flavor symmetries. The other model we consider is based on the coset SO(5) $\times$ U(1)$_P$ $\times$ U(1)$_X$/SO(4) $\times$ U(1)$_X$, in which the light state arises from an spontaneous breaking of a U(1)$_P$ symmetry of the new sector, that can also generate anomalous contributions to the couplings with gluons and photons. However at $1\sigma$ level the coupling of the bottom-quark demands U(1) charges of new composite fermions $\sim {\cal O}(100)$, introducing some tension. At $2\sigma$ level this requirement is relaxed, since the $\tau^+\tau^-$ channel with initial $b$-quarks is compatible with the SM. 

Besides the introduction of flavor symmetries, it is also interesting to consider other realizations of flavor, as in Ref.~\cite{Panico:2016ull}. In this case the Right-handed mixing angles are model dependent, and can be very small. As a consequence, for the SO(6)/SO(5) model the bounds from meson mixing can be fully relaxed. On the other hand, for the SO(5) $\times$ U(1)$_P$/SO(4) model, there are flavor violating contributions generated by the Left-handed mixing angles that are of CKM size, thus departures form universality of U(1)$_P$ Left-handed charges are constrained as in the anarchic case.

Last, it would be interesting to study models with a pseudoscalar arising from other cosets, in particular with unified groups, where one could expect to obtain contributions to the gluon coupling from the anomaly, as well as models with elementary weakly-coupled states.

\section*{Acknowledgments}
The authors would like to thank Ezequiel \'Alvarez for help with MGME and V\'{\i}ctor Mart\'{\i}n-Lozano for collaboration in the beginning of this article and for fruitful discussions.
The work of EA is partially supported by the ``Atracci\'on de Talento'' program (Modalidad 1) of the Comunidad de Madrid (Spain) under the grant number 2019-T1/TIC-14019 and by the Spanish Research Agency (Agencia Estatal de Investigaci\'on) through the grant IFT Centro de Excelencia Severo Ochoa SEV-2016-0597. The work is also partially supported by CONICET and ANPCyT under projects PICT 2016-0164, PICT 2017-2751, PICT 2017-0802, PICT 2017-2765 and PICT 2018-03682.

\bibliographystyle{JHEP}
\bibliography{lit}

\end{document}